\documentclass[12pt,letterpaper]{article}
\usepackage[a4paper, total={7in, 10in}]{geometry}

\usepackage{graphicx}
\usepackage{helvet}
\usepackage{authblk}
\usepackage{hyperref}
\usepackage{amsmath} 
\usepackage{amssymb} 
\usepackage{orcidlink} 
\usepackage{commath}
\usepackage{tikz}
\usepackage{xcolor}
\usepackage{dcolumn}
\usepackage{bm}
\usepackage{hyperref}

\hypersetup{
    colorlinks=true,
    linkcolor=blue,
    citecolor=blue,
    filecolor=blue,
    urlcolor=blue
}
\usepackage[super,comma,sort&compress]  
   {natbib}

\makeatletter
\renewcommand{\maketitle}{\bgroup\setlength{\parindent}{0pt}
\begin{flushleft}
  \textbf{\@title}
  
  \@author
\end{flushleft}\egroup}
\makeatother

\title{Nonreciprocal interactions drive intermittent melting of Coulomb clusters}
\date{}

\author[1]{Zhicheng Shu}
\author[1]{Wei-Chih Li}
\author[2]{Wentao Yu}
\author[1,3,*]{Justin C. Burton} 

\affil[1]{Department of Physics, Emory University, Atlanta, GA 30322, USA}
\affil[2]{Applied Physics and Materials Science, California Institute of Technology, Pasadena, CA 91125, USA}

\affil[3]{Lead contact}
\affil[*]{Correspondence: justin.c.burton@emory.edu}

\begin{document}

\maketitle

\section*{SUMMARY}

Complex systems out of equilibrium often experience intermittent oscillations between quiescent and highly dynamic states. While intermittency is usually driven by stochastic noise or external forcing, energy can also be sourced from field-mediated interactions between particles, which are often nonreciprocal and effectively violate Newton's 3rd law. Here we demonstrate how nonreciprocal interactions produce intermittency in clusters of charged micron-sized particles confined in a plasma sheath. 
Using three-dimensional particle tracking, we observe that vertical oscillations, induced by a noisy plasma environment, are parametrically coupled to the horizontal vibrational modes. Experiments and simulations show that nonreciprocal interactions strongly amplify this coupling, injecting energy into the system. This mechanism triggers explosive melting transitions from an ordered cluster to an ergodic gas-like state, and leads to intermittent switching between states over long time. Overall, our work identifies nonreciprocal interactions as a key mechanism through which strongly-coupled finite systems transform interaction-mediated activity into dynamical nonequilibrium states.

\section*{KEYWORDS}

Nonreciprocal interactions, dusty plasma, emergent activity, intermittency

\section*{INTRODUCTION}

Understanding how systems organize and function far from equilibrium remains one of the central challenges in statistical and soft matter physics. Nonequilibrium systems typically break detailed balance, continuously extract energy from their surroundings, and sustain entropy production \cite{bowick2022symmetry-thermodynamics-and-topology-in-active-matter, battle2016broken-detailed-balance}. As a result, they can exhibit dynamics and functionalities that are inaccessible in equilibrium. Nonequilibrium systems are ubiquitous in nature, spanning length scales from molecular motors to animal flocks \cite{schliwa2003molecular, julicher1997modeling, ballerini2008flock-animal-collective-behavior-depends-on-topological-rather-than-metric-distance, cavagna2010scale-Scale-free-correlations-flocks, nagy2010hierarchical-group-dynamics-in-pigeon-flocks, Ginelli2015Sheep}. Similar principles are increasingly exploited in the laboratory to design synthetic materials, including colloidal assemblies that self-organize into dynamical structures \cite{palacci2013Living-crystals-of-light-activated-colloidal-surfers, aubret2018targeted-assembly-and-synchronization-of-self-spinning-microgears,bricard2013emergence, bricard2015emergent, lefranc2025synthetic, maity2023spontaneous-demixing-of-binary-colloidal-flocks, osat2023non-reciprocal-multifarious-dself-organization, osat2024escaping-kinetic-traps-using-nonreciprocal-interactions} and active materials composed of work-generating units that support self-excited oscillations and locomotion \cite{scheibner2020-Odd-elasticity, baconnier2022Selective-and-collective-actuation-in-active-solids, veenstra2025adaptive-locomotion-of-active-solids, brandenbourger2019Non-reciprocal-robotic-metamaterials, veenstra2024Non-reciprocal-topological-solitons, veenstra2025nonreciprocal-breathing-solitons}.

A many-body system of interacting particles can be driven far from equilibrium by making each particle active. Here, ``activity'' typically means that individual constituents internally consume energy to generate motion \cite{marchetti2013hydrodynamics, bowick2022symmetry-thermodynamics-and-topology-in-active-matter, Active-particles-rmp}. In nature, microswimmers such as \textit{E. coli} convert chemical energy from ATP hydrolysis into self-propulsion \cite{berg1972chemotaxis-in-Escherichia-coli-analysed-by-three-dimensional-tracking, Active-particles-rmp}. Nonliving particles can acquire autonomous motion when energy is supplied externally through phoretic mechanisms such as light activation \cite{palacci2013Living-crystals-of-light-activated-colloidal-surfers, aubret2018targeted-assembly-and-synchronization-of-self-spinning-microgears}, chemical reactivity \cite{paxton2004catalytic-nanomotors, golestanian2005propulsion}, or DC electric fields \cite{bricard2013emergence,bricard2015emergent,lefranc2025synthetic}. Collections of such self-propelled units are referred to as active matter, which exhibits emergent behavior arising from their persistent energy consumption at the single-particle level, while the interactions between particles can be pairwise and reciprocal \cite{marchetti2013hydrodynamics}.

Even without single-particle activity, interacting particles can exhibit rich collective behavior when the interactions themselves become nonreciprocal, i.e., $\bm{F}_{ij}\neq -\bm{F}_{ji}$, effectively violating Newton's 3rd law \cite{ivlev2015statistical,loos2020irreversibility}. Such nonreciprocity naturally arises when forces are mediated through a driven environment. For example, wave-mediated interactions can support emergent collective activity, allowing two or more otherwise passive particles to harness energy from scattered waves and move coherently \cite{EllaKing2025Scattered_waves_fuel_emergent_activity, Morrell2025Nonreciprocal_wave-mediated_interactions_power_a_classical_time_crystal,wu2023hydrodynamic,BradyWu2025A_three_body_problem}. This ``social activity'' depends strongly on the spatial configuration of particles, and will generally involve non-pairwise interactions \cite{BradyWu2025A_three_body_problem,st2023dynamics, parker2025symmetry}. In fact, many living systems that exhibit flocking, schooling, and other coordinated motion involve both single particle activity and nonreciprocal, field-mediated interactions \cite{marchetti2013hydrodynamics,Oza2019PRX_Lattices_Flapping_Swimmers, Filella2018PRL_Collective_Fish_Behavior_Hydrodynamic_Interactions, tan2022odd-living-chiral-crystals}. 

Significant progress has been made in understanding collective behavior from nonreciprocity in the thermodynamic limit, such as the emergence of spatiotemporal oscillations and dynamical patterns that are inaccessible in equilibrium systems \cite{saha2020nonreciprocal-Cahn-Hilliard-model, you2020nonreciprocity-traveling-states, fruchart2021Non-reciprocal-phase-transitions, brauns2024nonreciprocal_pattern_formation,avni2025nonreciprocal_ising, nadolny2025Nonreciprocal_synchronization_of_active_quantum_spins, martin2025Transition-to-Collective-Motion-in-Nonreciprocal-Active-Matter, hickey2023nonreciprocalcilium,Guillet2025Melting_of_nonreciprocal_solids}. However, much less is known about how nonreciprocity affects the dynamics of finite systems with only a few coupled degrees of freedom where coarse-grained descriptions become insufficient. For such systems, analytical and numerical results have shown that nonreciprocity alone can break detailed balance and induce energy and information flow \cite{loos2020irreversibility, zhang2023entropy, ai2023brownian}. However, experimental data remains limited, particularly when steady states do not exist, and dynamics are complex. 

In this work, we experimentally demonstrate that nonreciprocal interactions between just two particles can generate unique nonequilibrium states marked by intermittent bursts of activity that drive transitions between quiescent and dynamical phases. Our experiments use a tractable model system: charged microparticles confined in a plasma sheath, commonly referred to as a dusty plasma. 
The levitated particles self-assemble into quasi two-dimensional (2D) crystalline clusters \cite{thomas1994plasma, chu1994direct, juan1998observation}, yet these clusters remain intrinsically far from equilibrium. Fluctuations in the surrounding plasma, such as variations in electric field and particle charge, introduce stochastic forcing that excites particle oscillations \cite{nunomura1999delaycharginginstability, ivlev2000influence, douglass2012determination, mendez2020origin,guga2017emergent, guga2020intermittent}. In addition, their plasma-mediated interactions are intrinsically nonreciprocal. When ions stream past a negatively charged particle, they form an asymmetric region of enhanced ion density below the particle (ion wake), leading to effective forces that violate action-reaction symmetry \cite{melzer1999transition, hebner2003dynamic, yu2025pnas, matthews2020dust}. There are several consequences of this nonreciprocity in dusty plasmas, \textcolor{black}{including mode coupling in extended crystals \cite{couedel2010direct, liu2010mode}, resonance instabilities in two or three-particle clusters \cite{kompaneets2006dust, yaroshenko2006vibrations, qiao2013mode,  lorin2019nonlinear, zhdanov2009mode}}, and temperature gradients in nonreciprocally coupled layers \cite{ivlev2015statistical}.

In our experiments, we observe intermittent transitions between an ordered Coulomb cluster and an ergodic, gas-like state \emph{without} external modulation. By tracking the full three-dimensional (3D) trajectories of all particles using laser-sheet tomography, we gain access to the complete phase space of the dynamics. Principal component analysis (PCA) of the particle motion reveals that melting is triggered by a parametric excitation of a horizontal breathing mode, which is driven by vertical oscillations of the cluster. The nonreciprocal interactions between the particles break the discrete time-translation symmetry of the vertical oscillation mode, manifesting as a subharmonic frequency component relative to the vertical confinement frequency. Although stochastic noise alone can excite vertical motion, we show using simulations that nonreciprocal interactions dramatically enhance the parametric mode-coupling by pumping the second harmonic of the breathing mode. This positive feedback loop drives an explosive growth in both horizontal and vertical modes that melts the cluster, which is not possible when the cluster is driven only by random noise. 

We use molecular dynamics simulations to further show that introducing nonreciprocal interactions induces intermittent dynamics characterized by abrupt melting events and bursts of kinetic energy, closely resembling the dynamics observed in experiments. In contrast, increasing noise merely raises the effective temperature uniformly without inducing intermittency marked by kinetic energy bursts. Together, these results demonstrate that our system offers a distinctive platform that links microscopic sources of nonreciprocity to the emergent dynamics of many-body systems. More broadly, we expect similar pathways of interaction-mediated activity to arise in other finite many-body systems and reshape their collective dynamics, from colloidal clusters to living cells to groups of animals.

\section*{RESULTS AND DISCUSSION}

\begin{figure*}[t]
    \centering
    \includegraphics[width=1\linewidth]{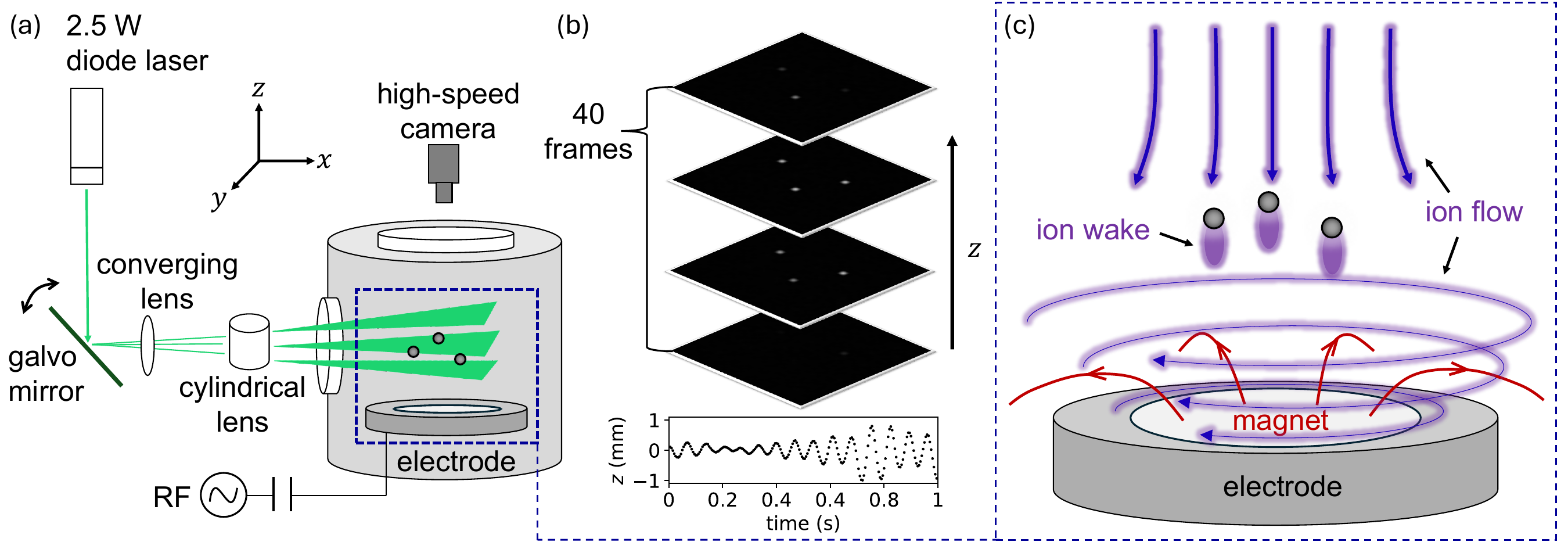}
    \caption{Sample preparation and imaging of dusty plasma clusters. (a) Schematic of the plasma chamber and the laser tomographic imaging. Micron-sized dust particles are dropped into the argon plasma discharge in the chamber. The high speed camera is synchronized with a scanning laser sheet which illuminates particles at different heights. (b) A stack of 40 frames of images is captured within one cycle of the laser scan, where particles at different $z$ positions are illuminated in different frames. \textcolor{black}{A sample time series from a single particle shows the vertical oscillations in $z$.}  (c) The charged dust particles are levitated above the electrode and confined within the center of the electrode due to the electric field gradient. The ions stream towards the electrode and form the ion wakes below the particles which make the interactions between particles effectively nonreciprocal. The down-streaming ions also drift in the azimuthal direction due to the magnetic field, which induces the cluster rotation. Note that the ion drift velocity is much smaller than the streaming velocity.}
    \label{fig1}
\end{figure*}

Figure~\ref{fig1}(a) shows the experimental setup of the dusty plasma and the tomographic imaging system. A radio frequency (RF) power supply operating at 13.56 MHz is connected to an aluminum electrode (diameter = 15 cm). A thin copper ring (outer diameter = 8.2 cm, inner diameter = 6.4 cm, thickness = 0.2 cm) is placed at the center of the electrode in order to generate the horizontal electrostatic confinement. A cylindrical neodymium magnet (diameter = 7.5 cm) is embedded inside the electrode and provides a nonuniform magnetic field ($B\approx$ 0.04 T) that enhances the electrostatic confinement and drives a vortical flow of ions. The RF power was adjusted between 2-10 W, which was sufficient to sustain an argon plasma inside the vacuum chamber. The argon pressure varied between 0.8-1.2 Pa depending on experimental conditions, and was maintained throughout each experiment to within 1\% using a feedback-controlled butterfly valve. Monodisperse melamine formaldehyde (MF) particles were released from a reservoir with a fine mesh at the top of the chamber. We used two batches of monodisperse particles with manufacturer-labeled diameters $10.55\pm 0.14$ $\mu$m and $12.8 \pm 0.32$ $\mu$m.

We tracked the motion of all particles in 3D using scanning laser sheet tomography \cite{yu20233d}. As shown in Fig.\ref{fig1}(a), an oscillating laser sheet scans the system along the $z$-direction at 100-500 Hz. A high-speed camera operating at 4-20 kHz ensures that each scan cycle produces a stack of 40 images at distinct $z$ positions (Fig.~\ref{fig1}(b)). We then track the particles and reconstruct their trajectories in 3D using a custom implementation of TrackPy \cite{Allan2021trackpy}. We apply a statistical method to reduce errors from pixel locking \cite{burov2017single}, especially in the $z$-direction where the resolution is limited by laser sheet width and scan frequency. Depending on the specific settings of camera resolution and laser scanning frequency, the ultimate resolution of each particle's position was 20-50 $\mu$m per pixel in the horizontal direction, and 50-100 $\mu$m per pixel in the $z$-direction. 

\subsection*{Forces in dusty plasma}

Particles experience several plasma-mediated forces, illustrated schematically in Fig.~\ref{fig1}(c). In the plasma, particles are bombarded by free electrons and ions and become negatively charged due to the higher electron temperature ($\approx 1.3$ eV \cite{mendez2020origin}). For the same reason, the electrode is also negatively charged, and forms a plasma sheath where the vertical electric field varies sharply in the $z$-direction. The electron and ion temperature and density also vary vertically in the sheath. Thus, the particle charge, $q(z)$, \textcolor{black}{typically of order $10^4$ electrons \cite{yu2025pnas}}, is position-dependent, which breaks translational symmetry in $z$. 
Particles are strongly confined at the edge of the plasma sheath, levitating approximately 1-2 cm above the electrode, where electrostatic repulsion balances gravity. In the horizontal direction, the electrode and copper ring provide a radial electric field that confines the particles. 

In the sheath region, ions stream downward to the negatively-charged electrode. The horizontal component of the magnetic field deflects the streaming ions and induces a vortical flow in the azimuthal direction. As a result, the particles experience a vortical ion-drag force that drives cluster rotation \cite{yu2025pnas}. The vertical component of the ion drift velocity is orders of magnitude larger than the azimuthal component. Ions are also deflected when they stream past a negatively-charged particle, and form a wake structure of enhanced positive charge beneath each particle \cite{vladimirov1995attraction, ishihara1997wake, lampe2000interactions, matthews2020dust}. The wake-mediated interaction is nonreciprocal: a particle sitting below another particle can be attracted to the wake of the upper one, whereas the upper particle is repelled \cite{melzer1999transition, hebner2003dynamic}. This nonreciprocity has been measured experimentally \cite{melzer1999transition, hebner2003dynamic}, and its functional dependence on particle separations was recently extracted from experimental data using machine learning \cite{yu2025pnas}. \textcolor{black}{While previous studies have shown that the ion wake-mediated attraction between particles can be weakened if the ions are magnetized \cite{carstensen2012ion, jung2018experiments}, the field strength in our experiments ($\sim$ 0.04 T) is well below that regime, thus we ignore the modification of wake structure due to the ion gyromotion.}

We summarize these interactions and (linearized) environmental forces in the following equations of motion. Let us denote the position of a particle $\bm{r}_i =(x_i, y_i, z_i)$ with $\bm{\rho}_i= (x_i, y_i)$, and the separation between two particles $\bm{r}_{ij}=\bm{r}_i-\bm{r}_j = \bm{\rho}_{ij}+ (z_i-z_j)\hat{\bf z}$, where $\bm{\rho}_{ij}=\rho_{ij}\hat{\bm{\rho}}_{ij}=(x_i-x_j,y_i-y_j)$. The equations of motion in the horizontal and vertical directions read: 
\begin{equation}
    \ddot{\bm{\rho}}_i = -\omega_h^2 \bm{\rho}_i + k_c^2  \hat{\bf z} \times \bm{\rho}_i    + \sum_{j\neq i} f^h_{ij}\hat{\bm{\rho}}_{ij}  - \gamma_i \dot{\bm{\rho}}_i
    \label{eom1}
\end{equation}
\begin{equation}
    \ddot{z_i} = -\omega_z^2 z_i +\sum_{j\neq i} f^z_{ij} - \gamma_i \dot{z_i}
    \label{eom2}
\end{equation}
where $\omega_h$ and $\omega_{z}$ are the horizontal and vertical confinement frequencies, and $k_c$ describes the vortical ion flow. Dotted variables refer to time differentiation. $f^h_{ij}$ and $f^z_{ij}$ are the horizontal and vertical components of the reduced interaction force $\bm{f}_{ij}(\bm{r}_{i},\bm{r}_{j}) = \bm{F}_{ij}(\bm{r}_{i},\bm{r}_{j})/m_i$, where $m_i$ is the mass of particle $i$. Finally, $\gamma_i$ is the damping coefficient, which depends on particle size. According to Epstein's law, for spherical MF particles with a density of 1510 kg$\cdot$m$^{-3}$ inside argon gas \cite{epstein1924resistance}, 
\begin{equation}
   \gamma_i = \frac{12.2P}{d_i} \mu\text{m}\cdot\text{Pa}^{-1}\cdot\text{s}^{-1}
    \label{epstein}
\end{equation}
where $d_i$ is the particle diameter and $P$ is the argon pressure.

Because the particle charge varies with $z$, $\omega_h$ and $F_{ij}$ are not invariant under translation in $z$, i.e., $\omega_h=\omega_h(z)$ and $\bm{F}_{ij}=\bm{F}_{ij}(\bm{\rho}_{ij},z_i,z_j)\propto q_i(z_i)q_j(z_j)$.  
Furthermore, the forces are nonreciprocal, i.e. $\bm{F}_{ij}\neq -\bm{F}_{ji}$, meaning that pairwise forces not only differ in magnitude, but also do not point along the displacement vector, $\bm{r}_{ij}$. Therefore, vertical oscillations are intrinsically coupled to horizontal oscillations even when particles have zero vertical separation \cite{kompaneets2006dust}, and this is the origin of parametric coupling that eventually melts the clusters.

Once entering the plasma, two or more particles settle into a quasi-2D rotating crystalline structure, where the horizontal confinement force is balanced by repulsive interaction forces, and the azimuthal ion drag force is balanced by damping. Even though the rotating crystal is the equilibrium ground state, we observe that rotating crystals are not always stable, but instead undergo intermittent melting transitions. In the following, we analyze fluctuations away from these force-balanced configurations, and uncover how small oscillations are amplified to trigger melting transitions through a combined influence of stochastic noise and crucially, nonreciprocal interactions. 

\subsection*{Intermittent melting of Coulomb clusters}

\begin{figure*}[t]
    \centering
    \includegraphics[width=1\linewidth]{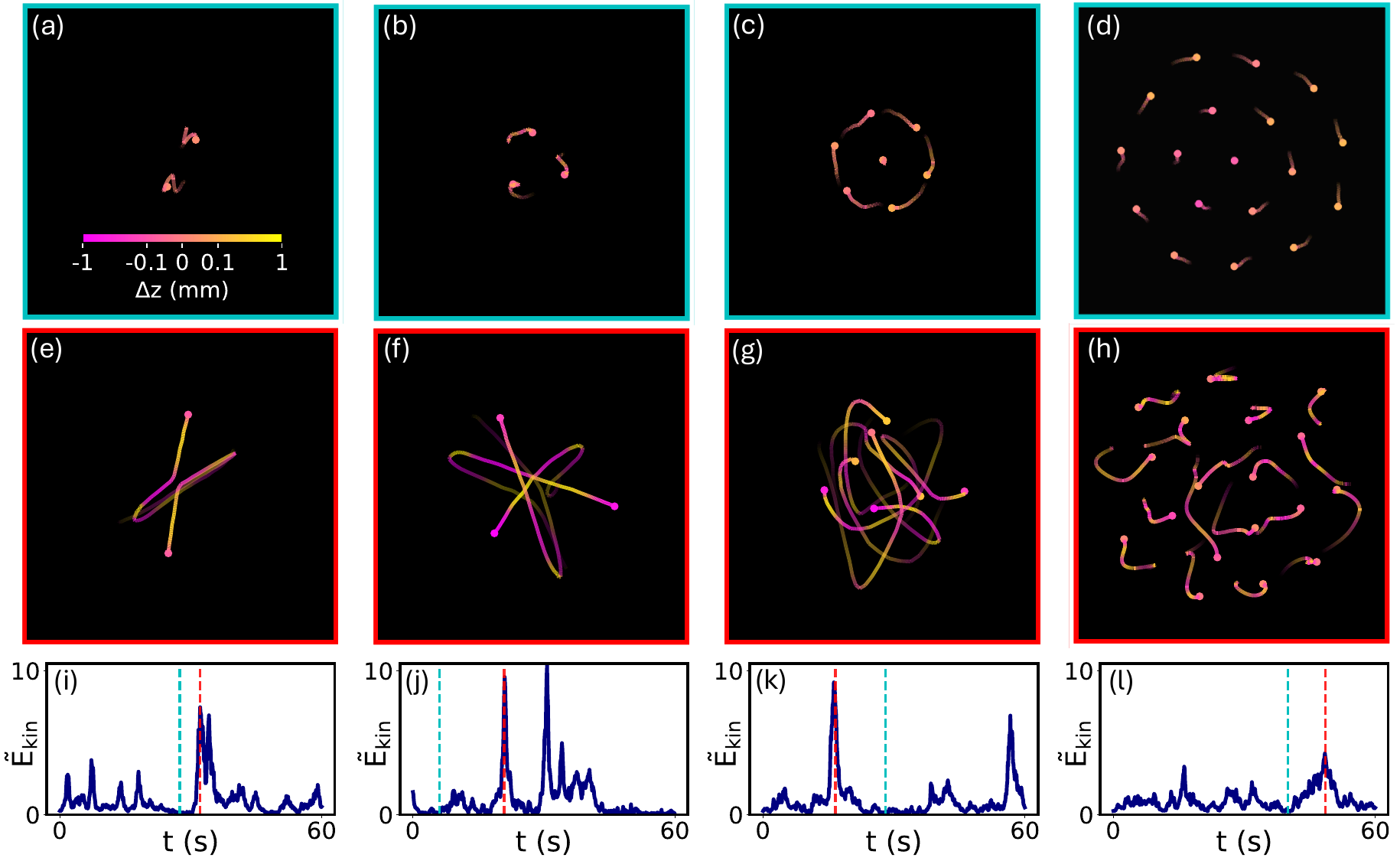}
    \caption{Intermittent melting of Coulomb clusters composed of 2, 3, 7, and 18 particles.  \textcolor{black}{The gas pressure for each experiment was 0.9, 0.9, 1.0, and 0.5 Pa, respectively.} (a-d) 3D trajectories of the quiescent crystalline state of the clusters. The trajectories are plotted in the $xy$-plane over 0.25 s. The size of each panel is 4 mm $\times$ 4 mm. The $z$-coordinate of each particle is represented by the color of the traces, where $\Delta z=0$ corresponds to the center of mass averaged over the whole time series. (e-h) 3D trajectories of the melted state of the clusters. (i-l) Total kinetic energy normalized by its time average over 60 s for each cluster. The blue (red) dashed line highlights the quiescent (melted) state shown above. }
    \label{fig2}
\end{figure*}

Figure~\ref{fig2} shows four Coulomb clusters observed in our experiments undergoing intermittent melting over 60 seconds (Videos S1-S8). The 3D trajectories of the particles are plotted for both the quiescent state (Fig.~\ref{fig2}(a-d)) and the melted state (Fig.~\ref{fig2}(e-h)).
The clusters stay in the quiescent state most of the time, where the motion is dominated by the uniform, rigid rotation of the cluster. The melting occurs intermittently, during which the kinetic energy increases dramatically. Surprisingly, the melting of clusters is not triggered by any externally applied modulations, but rather occurs spontaneously. In all cases, melting is initiated by the excitation of a horizontal oscillation mode. For clusters composed of 2, 3, and 7 particles,  all particles  oscillate radially with the same phase (except the center one in the 7-particle cluster). For the 18-particle cluster, an asymmetric breathing mode is excited, where particles at the inner shell and outer shell oscillate radially, but with opposite phase. \textcolor{black}{Although the 18-particle system exhibits melting transitions (Video S7), they are less explosive.}

The intermittent bursts of kinetic energy (Fig.~\ref{fig2}(i-l)) are hallmarks of the nonequilibrium dynamics exhibited by these clusters. The timescale between bursts, typically of order tens of seconds, is much longer than any microscopic timescale, suggesting that the intermittency arises from collective dynamics. \textcolor{black}{In much larger dusty plasma systems composed of hundreds of particles, similar intermittent dynamics manifest as switching between a highly-ordered crystalline state and an energetic gas \cite{guga2017emergent}. Here, the polydispersity of particle sizes is key to triggering the melting. Large or small particles create localized modes with a low participation ratio. These modes are the seeds of melting when the system is driven with vertical oscillations, yet nonreciprocity was not necessary for the intermittent dynamics \cite{guga2020intermittent}. Intermittency has also been observed in acoustically-levitated particle rafts \cite{wu2023hydrodynamic}, where scattered acoustic waves create $N$-body forces between particles that are also nonreciprocal. Energy is injected through local rearrangements of the raft's crystalline order, leading to energy bursts that must be dissipated through viscous damping.}

\textcolor{black}{In our experiments, intermittency is observed with only a few particles, and thus the mechanism of energy input from the environment is likely distinct from these previous examples.} To understand the emergence of this intermittency, we first quantify the cluster's motion in its quiescent state, characterizing its rotation and oscillation modes, and then examine how these modes evolve as kinetic energy bursts develop. \textcolor{black}{In the following, we focus our analysis on the three-particle cluster. This analysis applies for all clusters exhibiting explosive melting and a  symmetric breathing mode (2, 3, and 7 particles), while the 18-particle cluster will be discussed separately.} 



\subsection*{Three-particle cluster dynamics}

\begin{figure}
    \centering
    \includegraphics[width=0.5\linewidth]{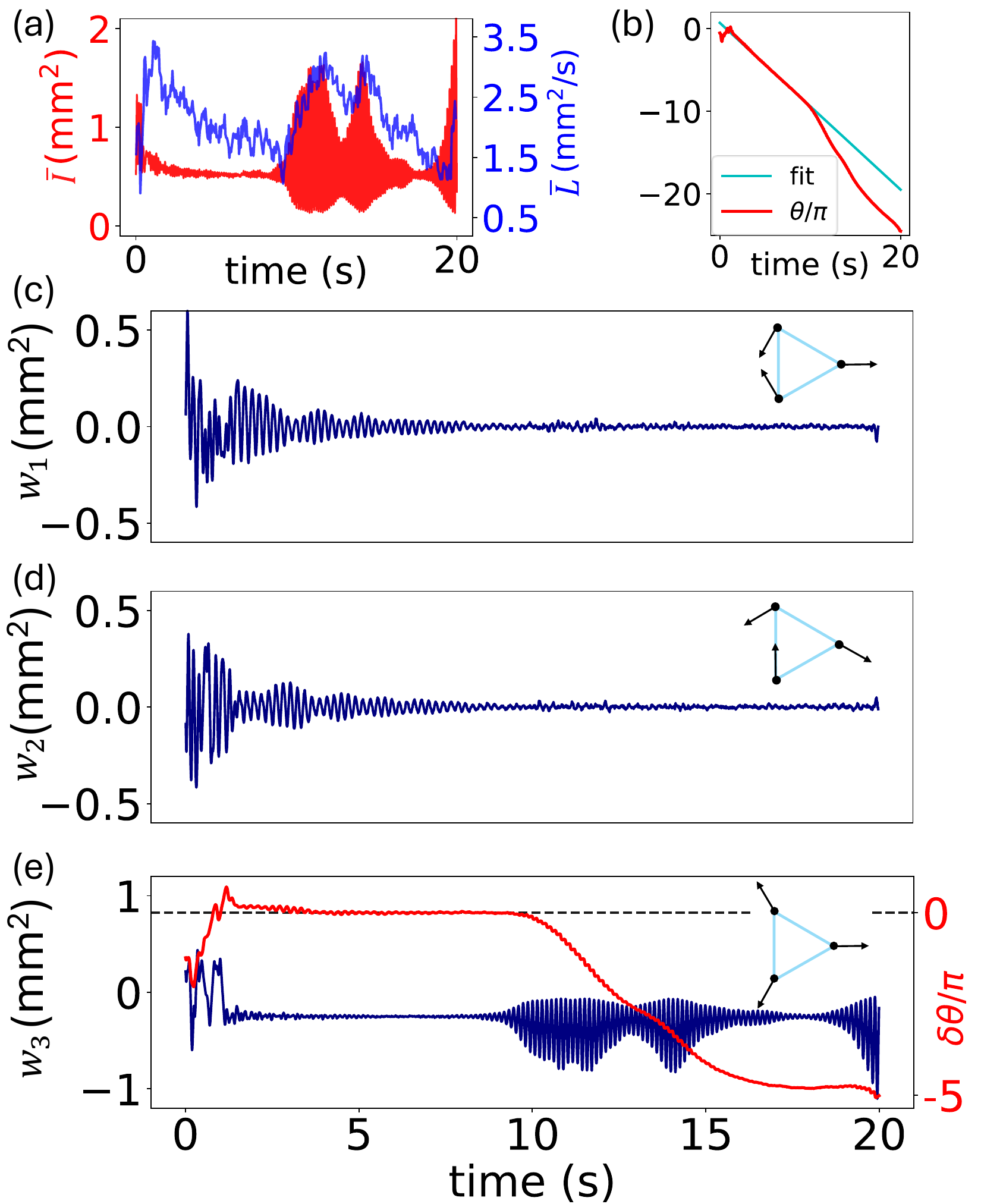}
    \caption{Rotation and oscillation modes of the three-particle cluster in experiments. 
    (a) Reduced moment of inertia $\bar{I}$, and reduced angular momentum $\bar{\mathbf{L}}$. Both $\bar{I}$ and $\bar{\mathbf{L}}$ increase when the breathing oscillation mode is excited. (b) Orientation angle of the cluster. The cyan line is a linear fit between $t=$ 4.5 s and $t=$ 9.5 s. 
    (c-e) Evolution of the $(w_1,w_2,w_3)$ variables. The insets show the corresponding displacements of the particles during the oscillation. The red data in (e) shows the deviation of $\theta$ from the linear fit in (b). }
    \label{fig3}
\end{figure}

In our experiments, the particles are nearly identical. However, they have slightly different masses that vary by a few percent and cannot be exactly determined. Thus we use the average manufacturer particle mass, $m_0$, for each sample to compute cluster-wide properties.  Figure~\ref{fig3}(a) shows the reduced moment of inertia, $I/m_0=\bar{I}=\sum_i |\bm{\rho}_i|^2$, and the reduced angular momentum, $\mathbf{L}/m_0=\bar{\mathbf{L}}=\sum_i  \bm{\rho}_i \times \dot{\bm{\rho}}_i$, for the first 20~seconds of the three-particle experiment highlighted in Fig.~\ref{fig2}(j). The oscillation of $\bar{I}$ arises from the variation of cluster size associated with the breathing mode excitation. As evident in Eq.~\ref{eom1}, $\bar{\mathbf{L}}$ is not conserved. If interactions are reciprocal, then the change of $\bar{\mathbf{L}}$ is determined by the torque due to the ion drag force and damping, leading to the following equation:
\begin{equation} \label{L-equation}
    \frac{d\bar{{L}}}{dt} = k_c^2\bar{I}-\gamma\bar{{L}}
\end{equation}
where $\bar{L}$ is the magnitude of the reduced angular momentum. Since we have assumed the particles are identical with mass $m_0$, they have the same damping rate according to Eq.~(\ref{epstein}). $\bar{\mathbf{L}}$ increases with $\bar{I}$ because particles experience a larger torque from ion drag force as they move farther away from the center of rotation~\cite{yu2025pnas,yu2022extracting}. On the other hand, the negative torque from damping force increases with $\bar{\mathbf{L}}$ as the cluster rotates faster. Eq.~(\ref{L-equation}) explains the initial relaxation of $\bar{\mathbf{L}}$ starting around $t=$ 2 s, and its increase as $\bar{I}$ increases. However, when the particles are not spaced symmetrically in the cluster, the nonreciprocal forces can lead to internal torques on the system, and Eq.~(\ref{L-equation}) becomes invalid.

Ignoring the center-of-mass translation, the dynamics of a three-particle cluster in two dimensions are fully described by four variables: the orientation angle, $\theta$, and three shape variables, $w_1$, $w_2$, and $w_3$~\cite{iwai1987geometric, efi2019self}. 
Here, $\theta$ is the angle between one edge of the triangular cluster and the $x$~axis. 
The shape variables capture the amplitudes of three oscillation modes of the cluster: two degenerate bending modes, $w_1$ and $w_2$, and the breathing mode, $w_3$. \textcolor{black}{ Defining $\bm{\rho}_{ij} = \bm{\rho}_i-\bm{\rho}_j$, we can write these modes explicitly: 
\begin{align}
w_1=&\frac{1}{6}\left(|\bm{\rho}_{13}|^2+|\bm{\rho}_{23}|^2-4\bm{\rho}_{13}\cdot\bm{\rho}_{23}\right),\\
w_2=&\frac{1}{2\sqrt{3}}\left(|\bm{\rho}_{13}|^2-|\bm{\rho}_{23}|^2\right),\\
w_3=&\frac{1}{\sqrt{3}}\left( \bm{\rho}_{13
} \times \bm{\rho}_{23}\right)\cdot\hat{\bf{z}}.
\label{w3}
\end{align}
The mode $w_1$ captures the symmetric bending where two angles of the triangle remain identical during the deformation, and mode $w_2$ captures the asymmetric bending where one side of the triangle extends while the other contracts.}
When there is no oscillation and the cluster is an equilateral triangle, $w_1=w_2=0$. 
Mode $w_3$ is the breathing mode, and its  
magnitude is proportional to the area of the triangle formed by the cluster, while its sign encodes the permutation parity of the three particles ($w_3$ switches sign whenever the ordering of the particles reverses). Therefore, switching between positive and negative values serves as a good indicator of cluster melting.

Figures~\ref{fig3}(b--e) show the evolution of the orientation angle and the shape variables for the first 20~seconds of the experiment. 
Recording starts at $t=0~\mathrm{s}$, immediately after a melting event, when the cluster was relaxing back to the quiescent state. 
During the quiescent state, $(w_1,w_2,w_3)$ return to their equilibrium values, and $\theta$ varies linearly in time. A deviation of $\theta(t)$ from a linear fit emerges at $t\approx10~\mathrm{s}$, marking the onset of the breathing mode excitation with a concomitant  increase in the oscillation of $w_3$. During the excitation, the deviation between $\theta(t)$ and its linear fit continues to grow, revealing that the cluster rotates faster when the breathing mode is excited. 
The accelerated rotation can be understood simply by considering conservation of angular momentum during a single oscillation with $\bar{\mathbf{L}}\neq0$, where small radial oscillations of a non-rigid body can cause global rotation. 
In contrast, if $\bar{\mathbf{L}}=0$, excitation of the breathing mode cannot induce rotation; instead, global rotation is driven by the excitation of $w_1$ and $w_2$ when they are out of phase~\cite{efi2019self}.

\subsection*{PCA extraction of oscillation modes}

\begin{figure*}[t]
    \centering
    \includegraphics[width=1\linewidth]{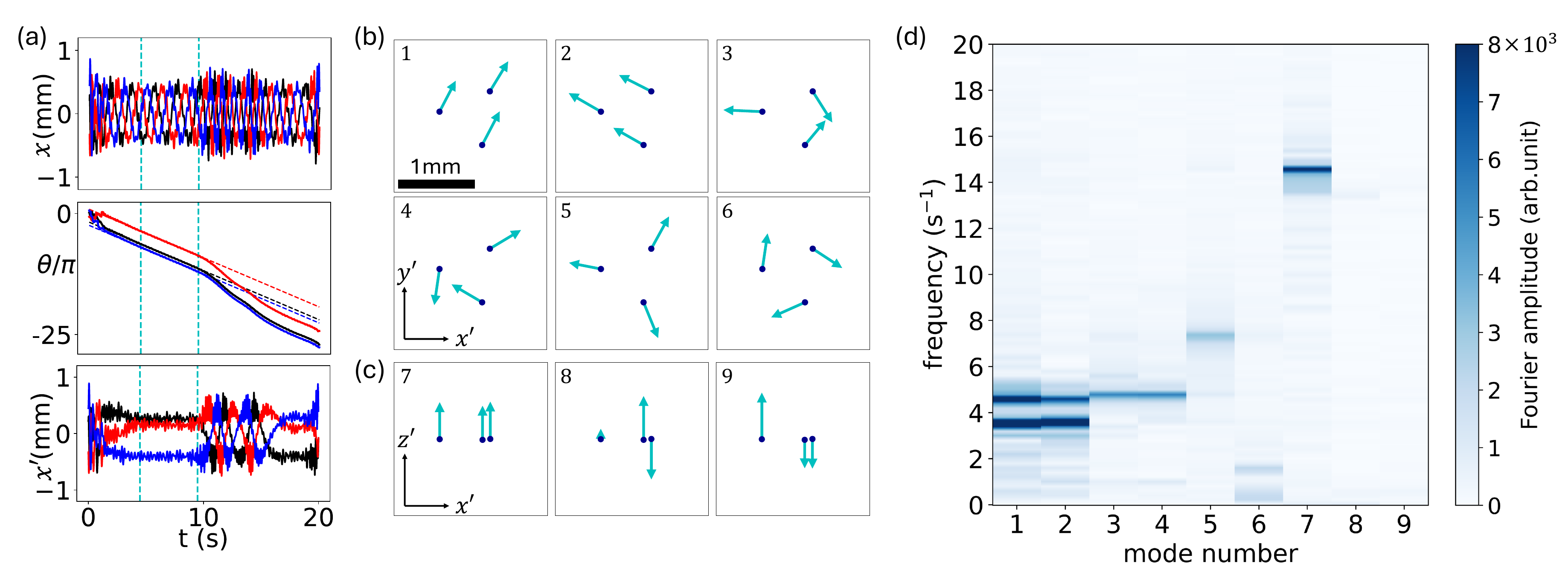}
    \caption{PCA mode analysis of the three-particle cluster experiment in the rotating reference frame. (a) The $x$-coordinate of each particle (red, blue, black) in the laboratory frame oscillates due to rotation. The rotation angle of each particle varies linearly until the breathing mode is excited. The 5 s region between the vertical cyan dashed lines denotes the quiescent state where the angular velocity of the rotating frame is computed. In the rotating frame ($x'$), the positions are fixed prior to the breathing mode excitation. 
    (b,c) Oscillation modes from PCA in the horizontal (1-6) and vertical (7-9) directions. Modes 1 and 2 are center-of-mass translations. Modes 3 and 4 are the two degenerate vibrating modes ($w_1$ and $w_2$). Mode 5 is the breathing oscillation, and mode 6 is the zero-frequency rotational mode. Mode 7 is the $z$ center-of-mass oscillation, and modes 8-9 are asymmetric vertical oscillations. The length of the arrows denotes relative displacements of the particles.  
    (d) Fourier spectrum of the PCA modes. Modes 1 and 2 have split peaks due to the Coriolis force. The vertical fluctuations are dominated by mode 7, whose frequency is approximately twice the breathing mode 5.}
    \label{figure4}
\end{figure*}

While the dynamics of a three-particle cluster can be described compactly using the shape variables, a more general approach is required to extract oscillation modes in larger clusters. We therefore apply principal component analysis (PCA) as an unbiased method for identifying collective modes directly from experimental data \cite{chen2021data, ivanov2005melting, brunton_kutz_2022_sec1_5}, and to reveal parametric coupling between modes that lead to melting. The workflow is summarized in Fig.~\ref{figure4}. 

PCA extracts orthogonal modes from the covariance of particle position fluctuations around an equilibrium configuration. The first step is to obtain the equilibrium configuration by removing the global rotation (Fig.~\ref{figure4}(a)). We first estimate the angular velocity, $\Omega$, of the rigid rotation during the quiescent state by measuring the angular velocities of individual particles about the center of mass, and taking the average. Particle coordinates are then transformed into the rotating frame via $\bm{\rho}'(t) = {\bf R}\cdot\bm{\rho}$ where ${\bf R}(\theta)$ is the rotation matrix about the $z$ axis, and $\theta = \Omega t$. In the rotating frame, Eq.~(\ref{eom1}) becomes:
\begin{equation} \label{EOM_rotating_frame}
    \begin{split}
        \ddot{\bm{\rho}}'_i &= -(\omega_h^2 - \Omega^2) \bm{\rho}'_i  - 2\Omega\hat{\bf z} \times \dot{\bm{\rho}}'_i + \sum_{j\neq i} f^h_{ij}\hat{\bm{\rho}}'_{ij}\\ &  - \gamma_i \dot{\bm{\rho}}'_i - (\gamma_i\Omega-k_c^2)\hat{\bf z} \times \bm{\rho}'_i.
    \end{split}
\end{equation}
This transformation has a few consequences. The effective confinement potential is reduced due to the centrifugal force, and the Coriolis force is non-negligible. The vortical force disappears when damping is balanced with the ion drag force. In this frame, particles fluctuate around fixed equilibrium positions without rotation when the cluster is in the quiescent state (Video S9).

We construct separate covariance matrices for the fluctuations of the horizontal and vertical coordinates in the rotating frame (Note S2). The eigenvectors of these matrices decompose the positional fluctuations into different oscillation modes, and the eigenvalues correspond to the contribution of each mode to the fluctuations about equilibrium. For the three-particle cluster, PCA identifies 6 horizontal modes, illustrated in Fig.~\ref{figure4}(b). There are two center-of-mass translational modes, two bending modes, the breathing mode, and the rotational mode. In addition, there are three vertical modes shown in Fig.~\ref{figure4}(c). Particle trajectories are projected onto each mode to obtain the corresponding time series of mode amplitudes, and the mode spectra are obtained by performing Fourier transforms on each time series, as shown in  Fig.~\ref{figure4}(d). The two translation modes are degenerate and have two peaks that correspond to $\omega_h\pm\Omega$. This frequency splitting results directly from the Coriolis force in the rotating reference frame (Note S1). The two bending modes are also degenerate as expected, while the breathing mode is the highest frequency horizontal mode. The bending modes and the breathing mode agree well with those described by the $(w_1,w_2,w_3)$ variables. Additionally, in the vertical direction, the particle motion is dominated by the center-of-mass translation mode.

This analysis demonstrates that PCA can serve as a reliable method of extracting oscillation modes from particle trajectories measured in experiments. While here we emphasize the three-particle system for simplicity, PCA works equally well on larger clusters, for example, 7 particles (Figs.~S1 and S2) and 18 particles (Figs.~S3 and S4).

\subsection*{Parametric coupling drives melting}

\begin{figure*}[t]
    \centering
    \includegraphics[width=1.0\linewidth]{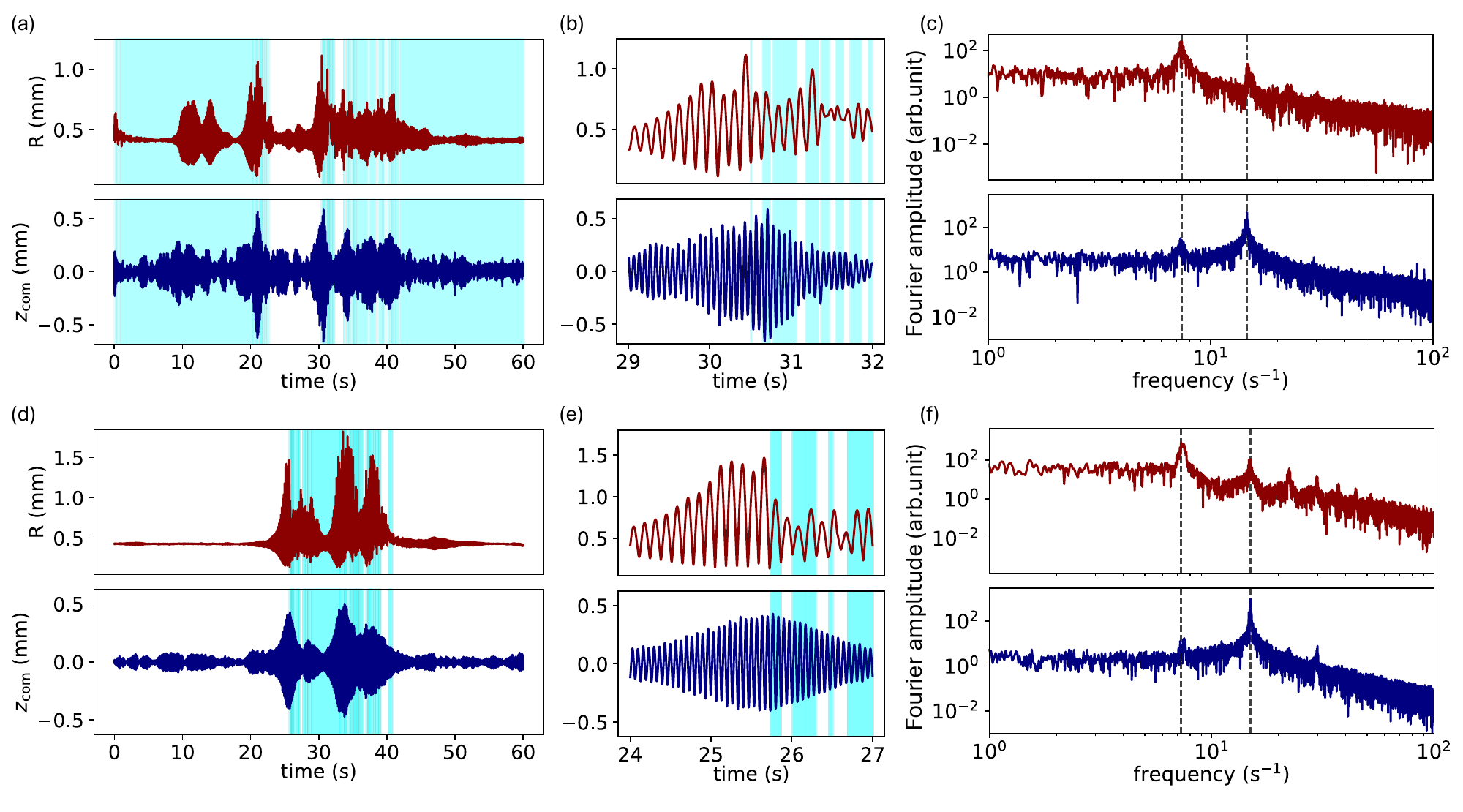}
    \caption{Simultaneous growth of the horizontal breathing mode and vertical center-of-mass mode drives melting in both experiment ($\text{a}$-$\text{c}$) and simulation ($\text{d}$-$\text{f}$) with $\tilde{q}=0.4, \sigma=0.1$. ($\text{a}$,$\text{d}$) Cluster radius, $R$, and the $z$ center of mass, $z_{\text{com}}$, of the three-particle cluster. The blue shaded regions indicate $w_3<0$. ($\text{b}$,$\text{e}$) $R$ and $z_{\text{com}}$ of the three-particle cluster near the onset of melting. The breathing mode amplitude increases until the cluster melts and ${w}_3$ begins to switch sign. The amplitude of the $z$ center-of-mass oscillation grows with the breathing mode. ($\text{c}$,$\text{f}$) Fourier spectrum of ${R}$ and ${z}_{\text{com}}$. The two dashed lines mark the peak frequencies of $f_b=7.42$ s$^{-1}$ ($7.28$ s$^{-1}$), and $f_z=14.57$ s$^{-1}$ ($14.95$ s$^{-1}$) in experiment (simulation).}
    \label{fig5}
\end{figure*}

Among all modes, two play a central role in the melting dynamics: the horizontal breathing mode and the vertical center-of-mass mode. In Fig.~\ref{fig5}(a), we plot the average cluster radius, $R=\sum_i\sqrt{x_i^2+y_i^2}/3$, which quantitatively indicates the breathing mode amplitude, and the vertical center-of-mass position $z_\text{com}$. Transitions between shaded and unshaded regions indicate where $w_3$ switches sign, and melting corresponds to densely-spaced transitions. In the melted state, both $R$ and $z_\text{com}$ exhibit large-amplitude fluctuations, and increase simultaneously before melting occurs (Fig.~\ref{fig5}(b)). In Fig.~\ref{fig5}(c), spectral analysis shows that the breathing mode has a peak frequency $f_b=7.42$ s$^{-1}$, while the vertical center-of-mass mode has a peak frequency $f_z=14.57$ s$^{-1}$, nearly twice the breathing mode frequency. This 2:1 frequency ratio is indicative of a parametric coupling: the breathing mode is parametrically modulated by the vertical oscillation. Such parametric coupling is consistently observed in all experiments where melting occurs, and is absent in experiments where the cluster remains quiescent (Figs.~S5 and S6). 

\textcolor{black}{The hybridization  of horizontal and vertical oscillation modes in a dusty plasma has previously been observed experimentally \cite{qiao2014mode} and predicted from theory using two and three particles \cite{kompaneets2006dust,qiao2013mode,yaroshenko2006vibrations}. This can occur when the horizontal and vertical modes are naturally close in frequency. However, the mode coupling observed in our system is of a parametric nature and drives explosive melting events.} Parametric instabilities often exist in periodically modulated systems. Common examples include Faraday waves: vertically shaking a flat fluid interface at frequency $\omega_D$ can excite standing surface waves with frequency $\omega_D/2$ \cite{miles1990parametrically}. The instability can be modeled by an oscillator coordinate $q$ whose fundamental frequency, $\omega_0$, is periodically modulated at frequency $\omega_D$, resulting in a damped Mathieu equation:
\begin{equation}
    \ddot{q} + \omega_0^2[1+\epsilon \cos(\omega_Dt)]q + \gamma\dot{q}= 0,
    \label{eq2}
\end{equation}
where $\gamma$ is the damping rate. When $\omega_D$ is close to $2\omega_0/n$ ($n$ is a positive integer), the solution will grow exponentially when the coupling strength $\epsilon$ exceeds a threshold \cite{kovacic2018mathieu}. 
In our Coulomb clusters, the parametric coupling between horizontal and vertical oscillation modes originates from the particles' charge dependence on vertical position in the plasma sheath \cite{nunomura1999delaycharginginstability, ivlev2000influence, douglass2012determination, mendez2020origin}. 
The equilibrium charge of a particle depends on the local electron and ion density, and local plasma potential, all of which vary sharply in $z$. The oscillation modes of a cluster are determined by the dynamical matrix, which includes interactions between the particles, and the interactions depend on the charge of each particle. Therefore, the frequencies of the horizontal oscillation modes of the cluster are modulated by the charge variation as the cluster oscillates vertically. 

Parametric instabilities can be excited in dust clusters by driving the system with a sinusoidal AC signal applied at the electrode \cite{yaroshenko2002parametric}. In contrast, the clusters in our experiment are not subject to any external periodic driving. The particles spontaneously oscillate in the vertical direction with finite amplitude  less than 0.5 mm (Fig.~\ref{fig5}a), which is two orders of magnitude larger than expected from thermal Brownian motion. \textcolor{black}{Spontaneous vertical oscillations with large amplitudes can arise from a fluctuating plasma environment, fluctuating charge, and delayed charging \cite{mendez2020origin, ivlev2000influence, nunomura1999delaycharginginstability}.}
If the vertical oscillation amplitude is larger than the threshold for parametric growth, the horizontal mode can be excited. However, this mode amplification is difficult to sustain since nonlinearity typically alters the mode frequency and stops the mode from growing further. Yet, we observe explosive melting preceded by simultaneous growth of the breathing and vertical center-of-mass modes (Fig.~\ref{fig5}(b)). 

By breaking reciprocity, the sum of the interaction forces in the system can be nonzero and act as an effective driving force on the system's center of mass. This is evident in the Fourier spectrum of the vertical center-of-mass oscillation of the cluster (Fig.~\ref{fig5}(c)), where besides the peak vertical confinement frequency at $f_z=14.57$ s$^{-1}$, there is a subharmonic peak near $f_z/2$. This subharmonic peak breaks the discrete time-translational symmetry of the vertical oscillation due to confinement, as if the cluster is driven by an additional external force. This effective driving force originates from the nonreciprocal interactions between particles, and is crucial in enhancing the parametric instability and producing large amplitude oscillations that lead to melting.

\subsection*{Simulating nonreciprocal Coulomb clusters}




To further elucidate the role of nonreciprocity, we performed simulations of our Coulomb clusters using a custom molecular dynamics code \cite{yu2025pnas}. The details of the simulation are described in Note S3, and here we give a brief summary. \textcolor{black}{Each simulation consists of three nearly identical spherical particles with average diameter of 10.55 $\mu$m, and size variations consistent with the manufacturer-labeled size variation of particles used in experiments. However, we found similar results using particles with exactly identical sizes. } In the horizontal plane, the particles are confined in a harmonic trap mimicking the electric field in the experiment. The particles also experience a vortical ion drag force which induces the rotation. In the vertical direction, the particles are subject to a linearly varying electric field, and gravity. The electrostatic force is determined by the field strength multiplied by the particle charge.

In the simulation, the negative charge of a particle increases in magnitude as the particle moves downwards, \textcolor{black}{accounting for the charge variation due to the varying ion and electron densities within the plasma sheath \cite{mendez2020origin,douglass2012determination}}. \textcolor{black}{At equilibrium $z=0$, the particle carries around $3\times10^4$ electrons}. The particles themselves interact through a Yukawa potential with a Debye screening length $\lambda_D$ = 0.45 mm. This interaction is reciprocal, however, it does not include interactions with the ion wakes beneath each particle (Fig.~\ref{fig1}), which are nonreciprocal. We model the ion wake as a cloud of positive charge attached at a distance of 0.3$\lambda_D$ below each particle and moving with the particle as if the cloud is massless \cite{kryuchkov2020strange,yu2025pnas,ivlev2015statistical}. The amount of charge carried by the positive charge cloud is $\tilde{q} |q|$, where $q$ is the instantaneous charge of the particle, and $\tilde{q}$ is a dimensionless constant. \textcolor{black}{An example plot of constant electric potential energy contours from a particle and its wake is shown in Fig.~S7. To recover reciprocal interactions, $\tilde{q}$ can be set to zero. It is possible to reduce a particle's charge when it enters another particle's ion wake or when the horizontal separation between two particles is small (well within a Debye length)} \cite{matthews2020dust}. This effect is not incorporated in our simulation. Finally, uncorrelated Gaussian noise of standard deviation $\sigma$ is added in the vertical acceleration for each particle to roughly model the spontaneous vertical oscillations of the particles observed in experiments. 

The simulations show qualitative agreement with the experiments, as illustrated in Fig.~\ref{fig5}(d-f) (Video S10). The breathing mode is spontaneously excited, and its amplitude grows dramatically leading to cluster melting. When the breathing mode is excited, the vertical oscillation amplitude increases correspondingly, far exceeding the amplitude due to Gaussian noise. Most importantly, in the Fourier spectrum of $z_{\text{com}}$, a subharmonic peak exists near $f_z/2$, suggesting the breaking of discrete time translational symmetry of the vertical confinement. \textcolor{black}{The subharmonic peak is also visible in simulations of the seven-particle cluster (Fig.~S8). For comparison, we also performed simulations with purely reciprocal interactions ($\tilde{q}=0$, Fig.~S9).} In this case, the breathing mode excitation and cluster melting can still occur, but require unphysical amounts of stochastic noise, resulting in vertical oscillation amplitudes exceeding those observed in experiments. Importantly, the subharmonic peak in the spectrum of $z_{\text{com}}$ is absent.

\subsection*{Nonreciprocity-enhanced parametric pumping}

\begin{figure*}[t]
    \centering
    \includegraphics[width=0.975\linewidth]{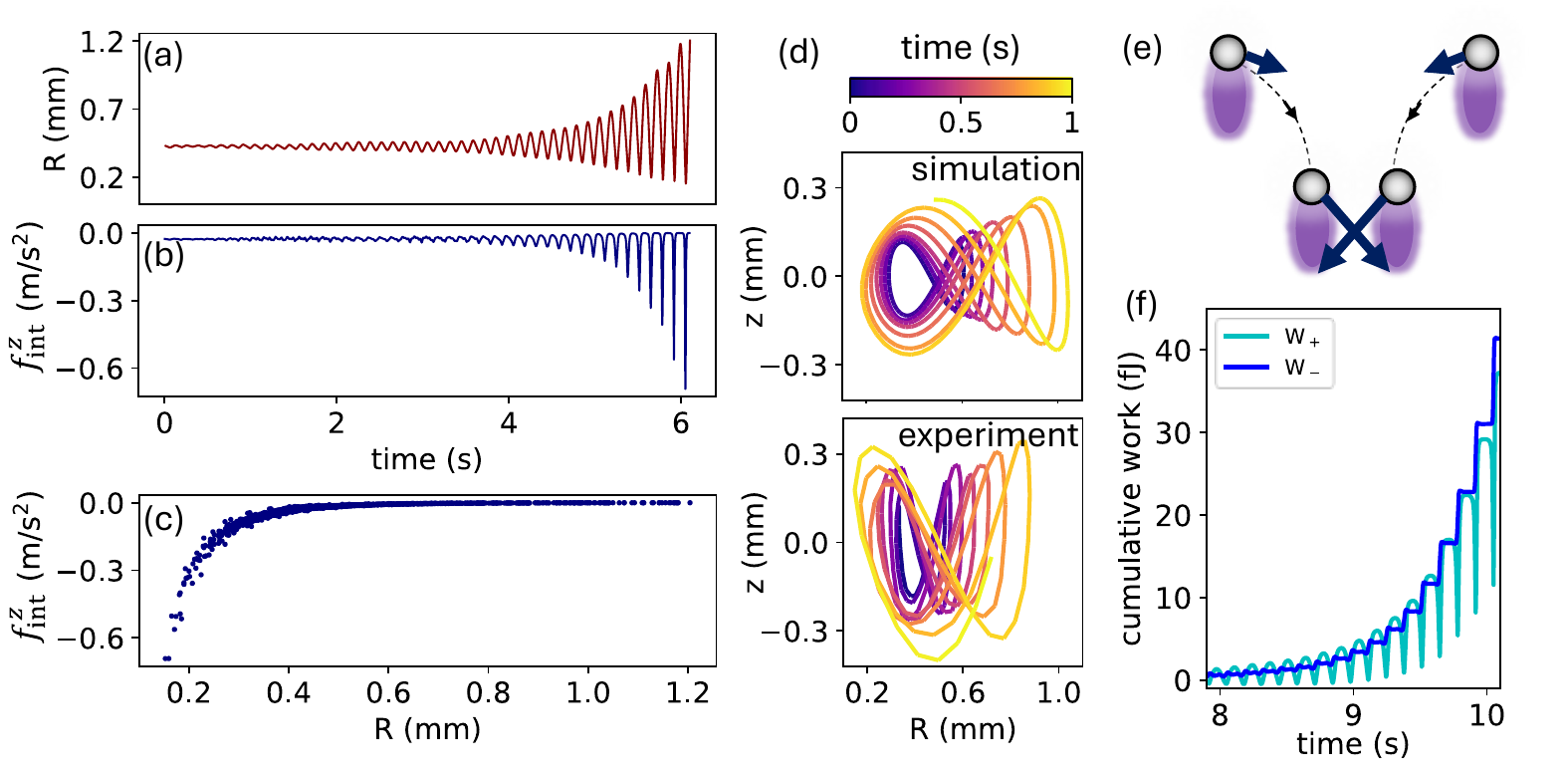}
    \caption{Nonreciprocal interactions drive vertical oscillations and enhance parametric pumping. ($\text{a}$-$\text{b}$) Cluster size ($R$) and $z$-component of the total interaction force ($f_{\text{int}}^z$) during excitation of the breathing mode in simulation with $\tilde{q}=0.4, \sigma=0.1$. \textcolor{black}{Note that the oscillations of $R$ and $f_{\text{int}}^z$ are in phase.} ($\text{c}$) $f_{\text{int}}^z$ as a function of $R$ during the same time period. 
    ($\text{d}$) Trajectories of $R$ and $z_{\text{com}}$ during a breathing mode excitation in both simulation and experiment. ($\text{e}$) Illustration showing that attraction to ion wakes in the $z$-direction drives the vertical oscillation during each cycle of the breathing mode. \textcolor{black}{The thick blue arrows represent the attraction between a dust particle and its neighboring particle's ion wake, which is enhanced as the two particles approach each other horizontally (the total force is still repulsive).} ($\text{f}$) Cumulative work $W_\pm$ done by the two components of the interaction forces, as defined in the text.}
    \label{fig6}
\end{figure*}

To elucidate how nonreciprocal interactions from ion wakes enhance the parametric instability, we examine the simulation immediately before a melting event. Figure~\ref{fig6}(a) shows the growth of $R$ during the excitation of the breathing mode. We calculate the total interaction force (divided by the average mass) acting on the cluster in the vertical direction during this process: $f_{\text{int}}^z=\sum_{ i\neq j}\bm{f}_{ij} \cdot \hat{\bf{z}}$. For reciprocal interactions, $f_{\text{int}}^z=0$ because $\bm
{f}_{ij}+\bm{f}_{ji}=0$ for any pair of identical particles. In contrast, as shown in Fig.~\ref{fig6}(b), $f_{\text{int}}^z$ is not zero, remains predominantly negative, and oscillates in phase with the breathing mode. Moreover, its magnitude increases substantially as the breathing mode amplitude grows. In Fig.~\ref{fig6}(c) we plot $f_{\text{int}}^z$ versus $R$. At small $R$, the vertical interaction force is strongly enhanced, whereas it approaches zero at large $R$. Thus, the total $z$-force on the system depends directly on the breathing mode amplitude.

We visualize the parametric pumping by plotting the trajectory of the three-particle cluster in the ($R$, $z_\text{com}$) phase space. Figure~\ref{fig6}(d) shows representative trajectories over a one-second interval prior to melting for both simulation and experiment. Owing to the parametric coupling between the two modes, the trajectories form a characteristic figure-eight shape whose size increases as the instability develops. Nonreciprocal interactions enhance this coupling, as sketched in Fig.~\ref{fig6}(e): when particles approach each other during the breathing motion, 
\textcolor{black}{the attraction between a particle and a neighboring particle's ion wake is significantly increased, and the vertical component of the interaction becomes more negative,} resulting in a larger net downward force acting on the cluster center of mass. Consequently, the breathing mode effectively drives vertical oscillations at $f_b$ through the nonreciprocal interaction force $f_{\text{int}}^z$. This is the origin of the subharmonic peak in the Fourier spectrum of $z_\text{com}$ (Fig.~\ref{fig5}(c) and \ref{fig5}(f)). 

\textcolor{black}{The subharmonic peak is visible for all clusters that exhibit a symmetric breathing mode (2, 3, and 7 particles), where melting is explosive since the large amplitude breathing mode also drives vertical oscillations. The asymmetric breathing mode observed in the 18-particle cluster also induces melting, but not by the same mechanism  shown in Fig.~\ref{fig6}(e). In a symmetric breathing mode, all particles can move closer to one another in phase, leading unequivocally to a net downward force on the center of mass. This is not true for the asymmetric breathing mode (mode 1, Fig. S3a). In experiments and simulations with reciprocal ($\tilde{q}=0$) and nonreciprocal ($\tilde{q}=0.4$) forces, we do not observe a subharmonic peak in the Fourier spectrum of the vertical center-of-mass. This comparison is shown in Fig.~S10(a-i). The net downward force on the center of mass, $f_\text{int}^z$, also does not vary with oscillation amplitude for the 18-particle cluster (Fig.~S10(j-k)).  While melting is observed, it is less energetic with smaller vertical oscillations. Thus, we conclude that nonreciprocity enhanced parametric pumping only occurs when the symmetric breathing mode is coupled to vertical oscillations by the mechanism depicted in Fig.~\ref{fig6}(e). }

For the three-particle cluster, our simulations illustrate how nonreciprocal interactions enhance the parametric instability by enabling a unidirectional channeling of energy into the system. We decompose the interaction force into two components: $\bm{F}_{+}^{ij}=\frac{1}{2}(\bm{F}_{ij}-\bm{F}_{ji})$ and $\bm{F}_{-}^{ij}=\frac{1}{2}(\bm{F}_{ij}+\bm{F}_{ji})$. For reciprocal interactions, $\bm{F}_{+}^{ij}=\bm{F}_{ij}$ and $\bm{F}_{-}^{ij}=0$. In Fig.~\ref{fig6}(f), we plot the cumulative work done by each component, $W_{\pm}(t)=\int_0^t\sum_{i\neq j}\bm{F}_{\pm}^{ij}\cdot\bm{v}_idt'$. Since $\bm{F}_{+}^{ij}$ is dominated by Coulomb repulsion, $W_+$ alternates between doing positive and negative work on the cluster during each breathing mode cycle, with small bias towards positive work due to the growth of the oscillation amplitude. In contrast, $W_-$ unidirectionally pumps energy into the cluster through the mechanism depicted in Fig.~\ref{fig6}(e). We note that for reciprocal interactions, $W_-=0$. Besides sourcing energy from the environment during each breathing mode cycle, we will also show that $f_{\text{int}}^z$ directly drives vertical oscillations near resonance at 2$f_b$, dramatically enhancing the growth of the breathing mode amplitude and melting the cluster.

\subsection*{A minimal model for nonreciprocal parametric pumping}

The mechanism by which nonreciprocal interactions enhance the growth of the breathing mode and induce cluster melting can be captured by a minimal model consisting of two parametrically coupled modes with nonreciprocal feedback. First, mode 1 is parametrically modulated by mode 2, such that the excitation of mode 1 depends on the amplitude of mode 2, as in conventional parametric coupling. Second, nonreciprocal interactions produce a driving force that depends on mode 1 and drives mode 2, thereby closing a positive feedback loop between the two modes. Consider two oscillators, $x_1$ and $x_2$ with natural frequencies $\omega_1$ and $\omega_2$, respectively, and damping rate $\gamma$. \textcolor{black}{In our experiments and simulations, $x_1$ represents the horizontal breathing mode, and $x_2$ represents the vertical center-of-mass translation mode.} The frequency $\omega_1$ is modulated by the oscillation of $x_2$ through the coupling parameter $\varepsilon$. $x_2$ is driven by a sinusoidal force on resonance and a feedback force $f$ which depends on $x_1$. The equations of motion read:
\begin{equation}
\begin{split}
    \ddot{x}_1 &+ \omega_1^2(1+\varepsilon x_2)x_1 + \gamma \dot{x}_1 = 0, \\
    \ddot{x}_2 &+ \omega_2^2 x_2 + \gamma \dot{x}_2 = a\cos(\omega_2 t)+f(x_1).
\end{split}
\label{minimal model}
\end{equation}
In reality, the vertical oscillation is driven by stochastic noise or other plasma-specific fluctuations, here we drive $x_2$ with a sinusoidal force for simplicity. 

In the absence of $f(x_1)$, $x_2$ oscillates with a fixed amplitude in steady state, $x_2(t) = a/\gamma\omega_2 \sin(\omega_2 t)$. When $\omega_2 \approx2\omega_1$, for large enough driving amplitude $a$, $x_1$ can grow exponentially. Assume $x_1$ oscillates at frequency $\omega_1$ with a slowly varying amplitude: $x_1=\frac{1}{2}A(t)e^{i{\omega}_1t}+\frac{1}{2}A^{*}(t)e^{-i{\omega}_1t}$. Then $A\propto e^{g_0 t}$, and the growth rate is given by :
\begin{equation}
    g_0 = -\frac{\gamma}{2}+\sqrt{\left(\frac{a {\varepsilon} {\omega}_1}{4\gamma\omega_2}\right)^2-\frac{\Delta^2}{4}}, \label{g0}
\end{equation}
where $\Delta=\omega_2-2\omega_1$. The feedback force $f(x_1)$ captures the nonzero total interaction force in the vertical direction, which depends on the breathing mode. \textcolor{black}{This reduced force can be measured in the simulations, and is shown in Fig.~\ref{fig6}(c) for three particles. The primary evidence for its effect in the experiments is the subharmonic peak in the vertical oscillation mode (Fig.~\ref{fig5}(c)).} Note that $f(0)=0$ because we subtract the value of $f_{\text{int}}^z$ at equilibrium $R$ so that $x_2=0$ represents the equilibrium position of the vertical center of mass of the cluster. Expanding $f(x_1)$ near the horizontal equilibrium position $x_1=0$ yields (Fig.~S11):
\begin{equation}
    f(x_1)=f_1x_1 + f_2x_1^2+\mathcal{O}(x_1^3).
\end{equation}
The linear term $f_1x_1$ generates a component of the feedback force oscillating near $\omega_1$, producing the subharmonic peak at $\omega_1$ in the vertical oscillation (Fig.~\ref{fig5}(f)). The quadratic term $f_2x_1^2$ plays a crucial role: it generates a component of the feedback force oscillating near $2\omega_1$, which drives $x_2$ near resonance. As a result, the amplitude of $x_2$ is no longer fixed by the external drive alone, but instead increases with the breathing mode amplitude, leading to a self-enhancing parametric instability. 

Writing $x_1=\frac{1}{2}A(t)e^{i{\omega}_1t}+\frac{1}{2}A^{*}(t)e^{-i{\omega}_1t}$ and $x_2=\frac{1}{2}B(t)e^{i{\omega}_2t}+\frac{1}{2}B^{*}(t)e^{-i{\omega}_2t}$, and applying the rotating wave approximation, we obtain the amplitude equations (Note S5):
\begin{subequations}
\begin{align}
\dot{A} &= -\frac{\gamma}{2}A - i\frac{\Delta}{2}A + \frac{i\varepsilon\omega_1}{4}B A^*
\label{A_amplitude} \\
\dot{B} &= -\frac{\gamma}{2}B + \frac{a}{2 i \omega_2} + \frac{f_2}{4 i \omega_2} A^2.
\label{B_amplitude}
\end{align}
\end{subequations}
Although the resulting amplitude equation is nonlinear and determining the full time evolution is challenging, it remains possible to define an instantaneous growth rate that captures the local amplification of the breathing mode amplitude: 
\begin{equation}
\begin{split}
    &g_{\text{inst}}=\frac{d}{dt}\ln{|A|} = \frac{1}{2}\frac{d}{dt}\ln{|A|}^2 =  \frac{1}{2|A|^2}\frac{d|A|^2}{dt} \\ &=\frac{1}{|A|^2}\text{Re}[\dot{A}A^*] 
    =-\frac{\gamma}{2}+\frac{\varepsilon\omega_1}{4}\text{Re}\left[iB\frac{(A^*)^2}{|A|^2}\right].
\end{split} \label{instant growth rate}
\end{equation}

To solve for $B$, we assume the following initial conditions: $|A(0)|\ll 1, \dot{A}(0)=0$ and $B(0)=a/(i\gamma\omega_2)$, i.e. at $t=0$, $x_2$ is in the steady state in response to the sinusoidal driving $a\cos{(\omega_2t)}$. Integrating Eq.\eqref{B_amplitude} gives 
\begin{equation}
    B(t) = B(0)+\frac{f_2}{4i\omega_2}e^{-\frac{\gamma}{2}t}\int_0^t e^{\frac{\gamma}{2}t'}A^2(t')dt'
    \label{B(t)}.
\end{equation}
In the usual parametric instability, $A(t)$ grows exponentially. Here, due to the feedback force $f(x_1)$, $A(t)$ is expected to grow with an increasing growth rate. Nevertheless, in order to evaluate Eq.\eqref{B(t)}, we assume $A(t)$ grows with a fixed bare growth rate $g_0$, i.e. $A(t) \approx A(0)\text{exp}({g_0t})$. Eq.\eqref{B(t)} then gives
\begin{equation}
\begin{split}
    B(t)&\approx B(0)+\frac{f_2A^2(0)}{4i\omega_2}e^{-\frac{\gamma}{2}t}\int_0^t e^{(2g_0+\frac{\gamma}{2})t'}dt'\\
    &=\frac{a}{i\gamma\omega_2}+\frac{f_2A^2(0)}{4i\omega_2(2g_0+\gamma/2)}\left(e^{2g_0t}-e^{-\frac{\gamma}{2}t}\right)\\
    &\approx \frac{a}{i\gamma\omega_2}+\frac{f_2A^2(t)}{4i\omega_2(2g_0+\gamma/2)}
    \label{B(t)_eff}.
\end{split}
\end{equation}
Plugging this into Eq.\eqref{instant growth rate} gives (Note S5)
\begin{equation}
    g_{\text{inst}} = g_0+\frac{\varepsilon\omega_1f_2|A|^2}{16\omega_2(2g_0+\gamma/2)}
    \label{g_inst}
\end{equation}

\begin{figure}
    \centering
    \includegraphics[width=0.5\linewidth]{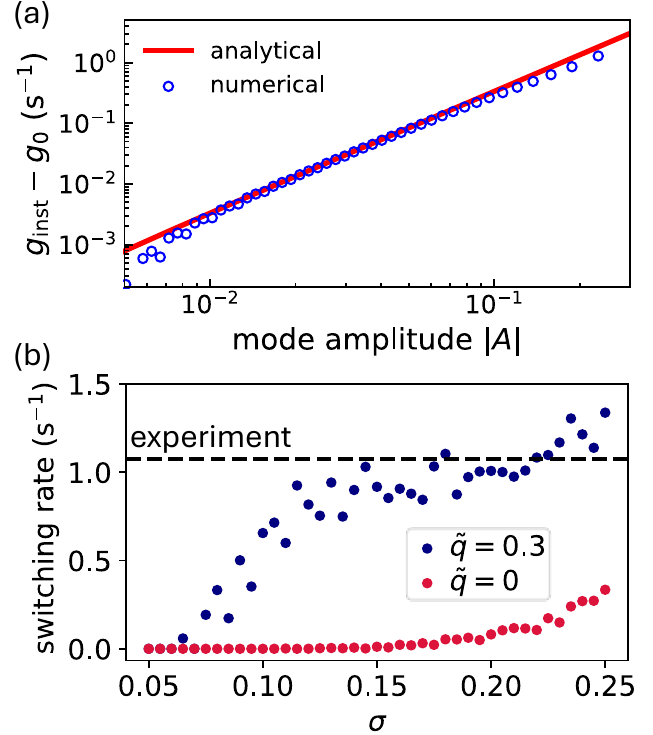}
    \caption{Nonreciprocity induces explosive mode growth and abrupt transitions to melting. (a) Instantaneous growth rate of $x_1$ (subtracting the bare growth rate $g_0=0.525$ s$^{-1}$) from the minimal model. Numerical results come from solving Eq.~\ref{minimal model} directly, and the analytical prediction is given in Eq.~\ref{g_inst}. (b) Switching rate of $w_3$ versus noise strength $\sigma$ for simulations with reciprocal ($\tilde{q}=0$) and nonreciprocal ($\tilde{q}=0.3$) interactions. The black dashed line represents the switching rate observed in the three-particle experiment (Fig.~\ref{fig2}j)}
    \label{fig7}
\end{figure}

Equation~\eqref{g_inst} highlights the central effect of nonreciprocal feedback---unlike the baseline parametric instability, where the growth rate is controlled by a fixed external drive (Fig.~S12), the parametric modulation here increases with the breathing-mode amplitude through the feedback force, $f_2x_1^2$. The instability is thus self-enhancing, leading to a substantially larger growth rate. In Fig.~\ref{fig7}(a), we numerically solve Eq.~\eqref{minimal model} and plot the instantaneous growth rate of $x_1$ as a function of its amplitude $|A|$ (the parameters used for the minimal model are listed in Note S6). The numerical results match the analytical prediction (Eq.~\eqref{g_inst}) exceedingly well and verify the quadratic dependence on $|A|$. In experiments, the explosive growth is cut off by the eventual melting of the Coulomb cluster (similar to simulations shown in Fig.~S13 and Fig.~S14). In the melted state, well-defined vibrational modes no longer exist, and drag forces from the neutral gas eventually allow for reformation of the cluster.

\subsection*{Nonreciprocity rewrites intermittency statistics}

We have shown that nonreciprocal interactions dramatically enhance the parametric pumping of the breathing mode and induce cluster melting. In the melted state, $w_3$ rapidly switches between negative and positive values (Eq.~(\ref{w3})). Therefore, its switching rate quantifies how often melting occurs. Figure~\ref{fig7}(b) shows the average switching rate of $w_3$, defined as the number of sign changes of $w_3$ divided by the total simulation (experiment) time, for simulations with reciprocal and nonreciprocal interactions at different noise strengths ($\sigma$). For reciprocal interactions, the switching rate increases gradually with $\sigma$. However, even for large noise ($\sigma = 0.25$), the switching rate remains significantly lower than the experimental value. In contrast, with nonreciprocal interactions, the switching rate increases much more rapidly with noise and approaches the experimental value at around $\sigma = 0.15$.

\begin{figure}
    \centering
    \includegraphics[width=0.5\linewidth]{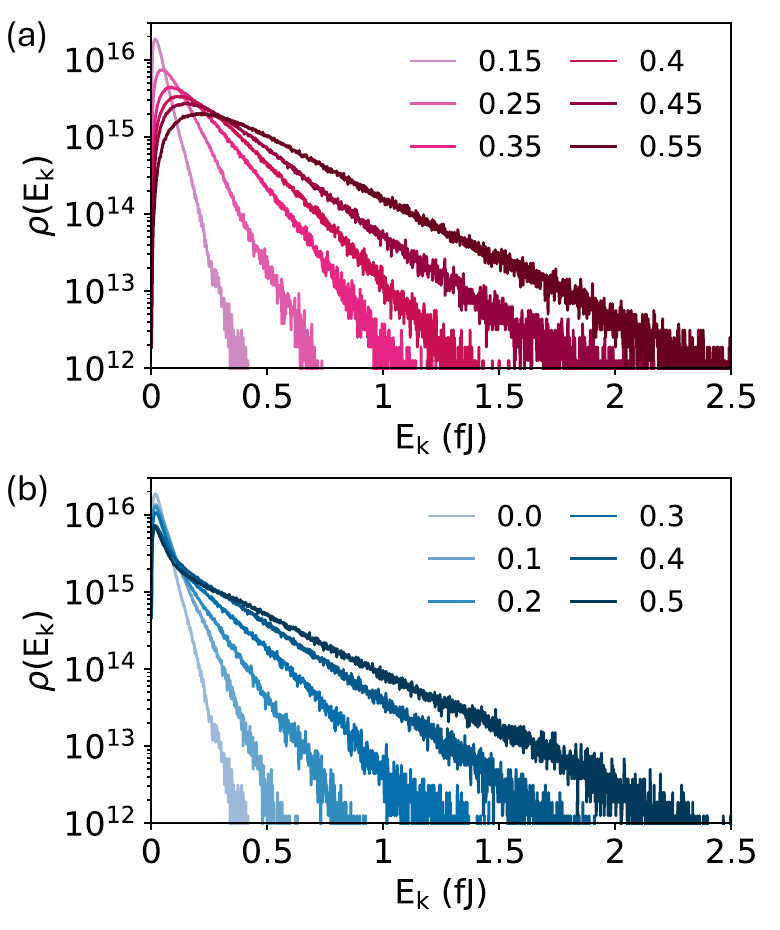}
    \caption{Stochastic noise and \textcolor{black}{nonreciprocal interactions} increase kinetic energy in different ways. (a) Probability distribution of kinetic energy for simulations with reciprocal interactions. Increasing the noise level ($\sigma$ is indicated in the legend) uniformly increases the temperature. (b) Probability distribution of kinetic energy for simulations at noise level $\sigma=0.15$. Increasing the \textcolor{black}{strength of ion wake ($\tilde{q}$ is indicated in the legend), which effectively enhances the nonreciprocal interactions,} results in two distinct regimes corresponding to quiescent and melted states.}
    \label{fig8}
\end{figure}

The parametric pumping relies on spatial symmetry of the cluster, where all particles move in unison (breathing mode). During the melted state, this mechanism no longer exists, kinetic energy is removed through damping, and the cluster re-forms. Over long times, the melting/recrystallization cycles are intermittent, and display distinct dynamics due to nonreciprocal interactions. To demonstrate this, we performed extensive simulations of three-particle clusters over a range of noise strength ($\sigma$) and \textcolor{black}{ion wake strength} ($\tilde{q}$). Figure~\ref{fig8} shows the probability distributions of the total cluster kinetic energy, $E_\text{k}$, obtained from these simulations. \textcolor{black}{Here, each simulation spans 2000 seconds, and the kinetic energy is measured 500 times per second.} In the absence of \textcolor{black}{nonreciprocal interactions}, increasing $\sigma$ simply thermalizes the system, leading to a higher effective temperature and a kinetic energy distribution consistent with a Maxwell-Boltzmann (MB) velocity distribution, $\rho(E_{\text{k}})\propto\sqrt{E_{\text{k}}}(k_BT)^{-3/2}\exp{(-E_{\text{k}}/k_BT)}$, where $k_B$ is the Boltzmann constant and $T$ is the effective temperature (Fig.~\ref{fig8}(a)). In contrast, for a fixed noise strength, increasing $\tilde{q}$ produces two distinct regimes: a peak at low energies (quiescent periods), and a high-energy tail corresponding to melted states (Fig.~\ref{fig8}(b)). This indicates strong deviations from equilibrium statistics.

\begin{figure}
    \centering
    \includegraphics[width=0.5\linewidth]{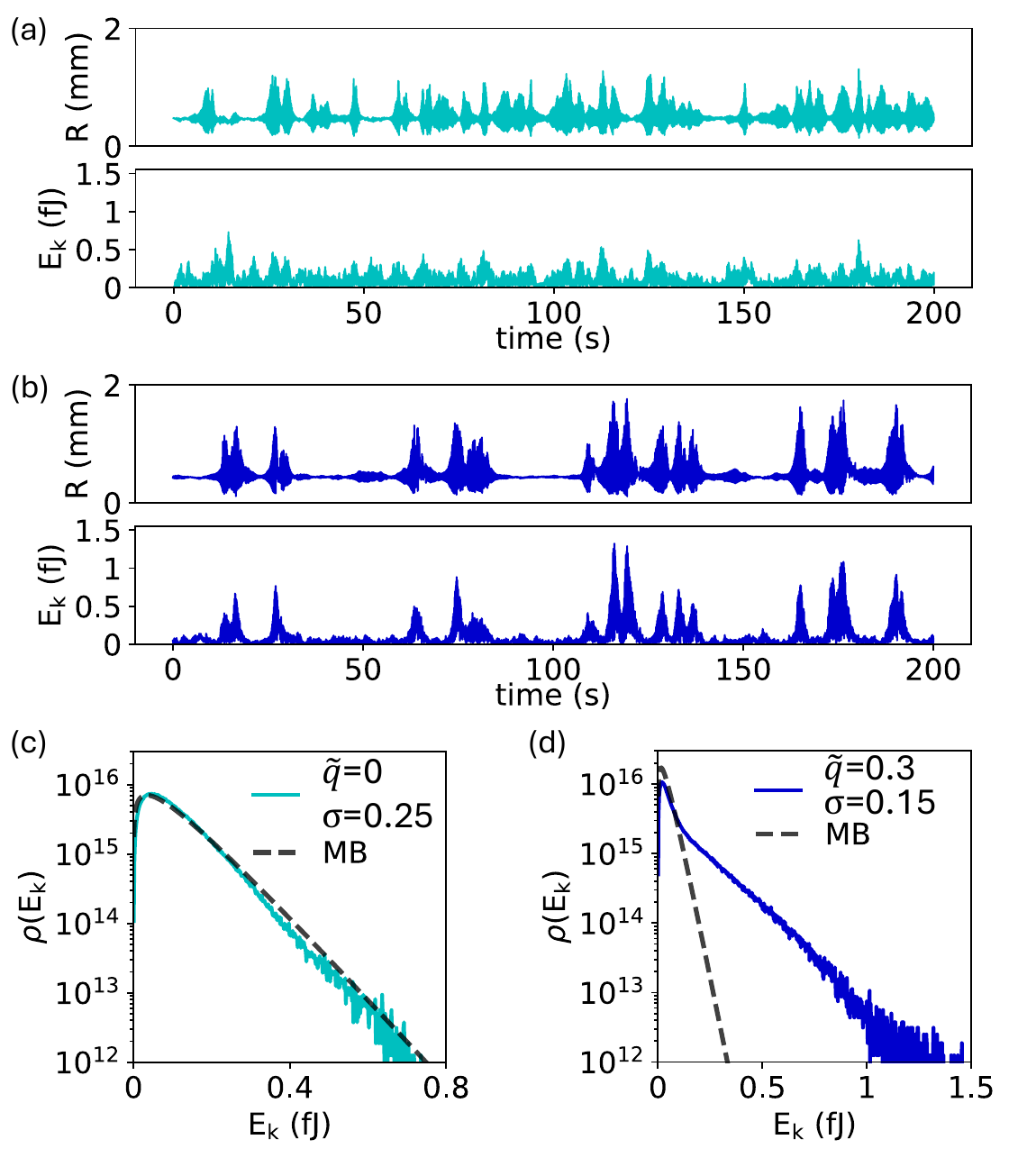}
    \caption{Nonreciprocal interactions induce punctuated bursts of kinetic energy. (a,b) Time series of average cluster radius and total kinetic energy from simulations with $\tilde{q}=0, \sigma=0.25$ (a) and $\tilde{q}=0.3, \sigma=0.15$ (b). (c) Kinetic energy probability distribution with noise only. The data is well-fit by a MB energy distribution with $T\approx5\times10^6$ K and $\langle E_{\text{k}}\rangle=$ 0.102 fJ. (d) Kinetic energy probability distribution with \textcolor{black}{nonreciprocal interactions}. The average kinetic energy is $\langle E_{\text{k}}\rangle=$ 0.118 fJ. The dashed line is an MB fit for a simulation with the same noise strength but $\tilde{q}=0$, giving $T\approx2\times10^6$ K. }
    \label{fig9}
\end{figure}

These distinct kinetic energy distributions reflect qualitatively different intermittent dynamics induced by nonreciprocal interactions. Figures~\ref{fig9}(a) and \ref{fig9}(b) show representative time series of $R$ and $E_{\text{k}}$ over 200 s for simulations with reciprocal and nonreciprocal interactions, respectively. The two simulations produce states with intermittent variations in $R$ and similar average kinetic energy, but $E_\text{k}$ is intermittent only when nonreciprocal interactions are present. With \textcolor{black}{nonreciprocal interactions}, the vertical center-of-mass mode increases concomitantly with the breathing mode (Fig.~\ref{fig5}(d)), whereas the vertical mode is roughly insensitive to the breathing mode with reciprocal interactions. Since the vertical mode carries a significant fraction of the kinetic energy, the system-wide dynamics are intermittent. Figure~\ref{fig9}(c-d) shows probability distributions of the kinetic energy in each case. In the absence of \textcolor{black}{nonreciprocal interactions}, the breathing mode is excited by large random noise. The fluctuation of kinetic energy is determined by the noise strength, and its distribution is well-fit by a MB energy distribution. However, even at substantially lower noise strength, nonreciprocal interactions facilitate a unidirectional injection of energy into the system during parametric coupling. This results in intermittent bursts of global kinetic energy that closely resemble those observed experimentally, and consequently a strongly nonequilibrium energy distribution with two regimes and high-energy tails. 

\section*{CONCLUSION}

We have experimentally demonstrated that nonreciprocal interactions strongly enhance the parametric instability between two oscillation modes in a Coulomb cluster, ultimately driving the cluster melting. Unlike conventional parametric pumping induced by external driving, where the instability is controlled by the driving strength, here nonreciprocal interactions act as an emergent internal driving mechanism that enables unidirectional energy injection into the cluster. As a result, the system enters an intermittent nonequilibrium state characterized by alternations between quiescent crystalline phases and melted, ergodic phases. Through simulations incorporating nonreciprocal interactions, we reproduce intermittent melting dynamics that closely resemble those observed in experiments, and confirm that such intermittency is absent when the system is driven solely by random noise. 

Beyond Coulomb clusters, our results identify a pathway by which strongly-coupled many-body systems can sustain dynamical nonequilibrium behavior through activity originating from microscopic nonreciprocity. Unlike conventional active matter, where energy is injected at the level of individual constituents, the activity here is mediated by nonreciprocal interactions and emerges collectively. Similar forms of interaction-mediated activity arise naturally in a variety of physical systems, including colloidal assemblies in acoustic and light fields \cite{EllaKing2025Scattered_waves_fuel_emergent_activity, BradyWu2025A_three_body_problem, st2023dynamics, parker2025symmetry}, and mechanical metamaterials \cite{veenstra2025adaptive-locomotion-of-active-solids}, where state manipulation through parametric pumping could benefit from such emergent activity \cite{mao2025structural, shi2025electrostatics}. Moreover, living systems often involve both single particle activity from energy consumption, and field-mediated interactions, i.e., hydrodynamics or chemical cues, that can facilitate collective motions such as flocking and schooling \cite{marchetti2013hydrodynamics}. How the interplay between activity and nonreciprocal (and possibly nonpairwise) interactions shapes collective behavior requires further exploration. Therefore, our results highlight how nonreciprocity-induced emergent activity--here realized through parametric coupling between two oscillation modes--can play an important role in a broad class of strongly-coupled, many-body systems.

\newpage

\section*{METHODS}

Details regarding the methods can be found in the supplemental information.


\section*{RESOURCE AVAILABILITY}


\subsection*{Lead contact}


Requests for further information and resources should be directed to and will be fulfilled by the lead contact, Justin C. Burton (justin.c.burton@emory.edu).

\subsection*{Materials availability}


This study did not generate new materials.

\subsection*{Data and code availability}


The particle-trajectory data from experiments and simulations, together with the associated analysis code, have been deposited at Zenodo: https://doi.org/10.5281/zenodo.21263442.

\section*{ACKNOWLEDGMENTS}


This material is based upon work supported by the NSF under award numbers 2010524 and 2409416, and by the Gordon and Betty Moore Foundation, Grant DOI 10.37807/gbmf12256.

\section*{AUTHOR CONTRIBUTIONS}
Z.S., W.-C.L., W.Y., and J.C.B. conceived the research and performed the experiments. Z.S. and J.C.B. performed the simulations and analyses. All authors contributed to writing of the manuscript.


\section*{DECLARATION OF INTERESTS}


The authors declare no competing interests.

\section*{DECLARATION OF GENERATIVE AI AND AI-ASSISTED TECHNOLOGIES}


During the preparation of this work, the authors used ChatGPT to assist with language polishing. The authors reviewed and edited the content as needed and take full responsibility for the content of the publication.

\section*{SUPPLEMENTAL INFORMATION}
Document S1. Notes S1-S6, Figures S1-S15, and Table S1.

\noindent Video S1. 2-particle melted state.

\noindent Video S2. 2-particle quiescent state.

\noindent Video S3. 3-particle melted state.

\noindent Video S4. 3-particle quiescent state.

\noindent Video S5. 7-particle melted state.

\noindent Video S6. 7-particle quiescent state.

\noindent Video S7. 18-particle melted state.

\noindent Video S8. 18-particle quiescent state.

\noindent Video S9. 3-particle quiescent state, in lab frame and the rotating reference frame. 

\noindent Video S10. 3-particle melted state, in both experiment and simulation ($\tilde{q}=0.4, \sigma=0.1$).

\noindent Video S11. Relaxation of 3-particle clusters from the melted state to the quiescent state, in both experiment and simulation ($\tilde{q}=0.4, \sigma=0.1$).

\bibliography{references}

@PREAMBLE{
 "\providecommand{\noopsort}[1]{}" 
 # "\providecommand{\singleletter}[1]{#1}%" 

}

@article{zhdanov2009mode,
  title={Mode-coupling instability of two-dimensional plasma crystals},
  author={Zhdanov, SK and Ivlev, AV and Morfill, GE},
  journal={Physics of Plasmas},
  volume={16},
  number={8},
  year={2009},
  publisher={AIP Publishing},
  url={https://pubs.aip.org/aip/pop/article-abstract/16/8/083706/316550/Mode-coupling-instability-of-two-dimensional?redirectedFrom=fulltext}
}

@article{yu2022extracting,
  title={Extracting forces from noisy dynamics in dusty plasmas},
  author={Yu, Wentao and Cho, Jonathan and Burton, Justin C},
  journal={Physical Review E},
  volume={106},
  number={3},
  pages={035303},
  year={2022},
  publisher={APS},
  url={https://journals.aps.org/pre/abstract/10.1103/PhysRevE.106.035303}
}

@article{kryuchkov2020strange,
  title={Strange attractors induced by melting in systems with nonreciprocal effective interactions},
  author={Kryuchkov, Nikita P and Mistryukova, Lukiya A and Sapelkin, Andrei V and Yurchenko, Stanislav O},
  journal={Physical Review E},
  volume={101},
  number={6},
  pages={063205},
  year={2020},
  publisher={APS},
  url={https://journals.aps.org/pre/abstract/10.1103/PhysRevE.101.063205}
}

@article{bricard2013emergence,
  title={Emergence of macroscopic directed motion in populations of motile colloids},
  author={Bricard, Antoine and Caussin, Jean-Baptiste and Desreumaux, Nicolas and Dauchot, Olivier and Bartolo, Denis},
  journal={Nature},
  volume={503},
  number={7474},
  pages={95--98},
  year={2013},
  publisher={Nature Publishing Group UK London},
  url={https://www.nature.com/articles/nature12673}
}

@article{bricard2015emergent,
  title={Emergent vortices in populations of colloidal rollers},
  author={Bricard, Antoine and Caussin, Jean-Baptiste and Das, Debasish and Savoie, Charles and Chikkadi, Vijayakumar and Shitara, Kyohei and Chepizhko, Oleksandr and Peruani, Fernando and Saintillan, David and Bartolo, Denis},
  journal={Nature communications},
  volume={6},
  number={1},
  pages={7470},
  year={2015},
  publisher={Nature Publishing Group UK London},
  url={https://www.nature.com/articles/ncomms8470}
}

@article{lefranc2025synthetic,
  title={Synthetic Quorum Sensing and Absorbing Phase Transitions in Colloidal Active Matter},
  author={Lefranc, Thibault and Dinelli, Alberto and Fern{\'a}ndez-Rico, Carla and Dullens, Roel PA and Tailleur, Julien and Bartolo, Denis},
  journal={Physical Review X},
  volume={15},
  number={3},
  pages={031050},
  year={2025},
  publisher={APS},
  url={https://journals.aps.org/prx/abstract/10.1103/8csn-71jk}
}

@article{st2023dynamics,
  title={Dynamics of acoustically bound particles},
  author={St. Clair, Nicholas and Davenport, Dominique and Kim, Arnold D and Kleckner, Dustin},
  journal={Physical Review Research},
  volume={5},
  number={1},
  pages={013051},
  year={2023},
  publisher={APS},
  url={https://journals.aps.org/prresearch/abstract/10.1103/PhysRevResearch.5.013051}
}

@article{marchetti2013hydrodynamics,
  title={Hydrodynamics of soft active matter},
  author={Marchetti, M Cristina and Joanny, Jean-Fran{\c{c}}ois and Ramaswamy, Sriram and Liverpool, Tanniemola B and Prost, Jacques and Rao, Madan and Simha, R Aditi},
  journal={Reviews of modern physics},
  volume={85},
  number={3},
  pages={1143--1189},
  year={2013},
  publisher={APS},
  url={https://journals.aps.org/rmp/abstract/10.1103/RevModPhys.85.1143}
}

@article{bowick2022symmetry-thermodynamics-and-topology-in-active-matter,
  title={Symmetry, thermodynamics, and topology in active matter},
  author={Bowick, Mark J and Fakhri, Nikta and Marchetti, M Cristina and Ramaswamy, Sriram},
  journal={Physical Review X},
  volume={12},
  number={1},
  pages={010501},
  year={2022},
  publisher={APS},
  url={https://journals.aps.org/prx/abstract/10.1103/PhysRevX.12.010501}
}

@article{Ginelli2015Sheep,
  title     = {Intermittent collective dynamics emerge from conflicting imperatives in sheep herds},
  author    = {Ginelli, Francesco and Peruani, Fernando and Pillot, Marie-Hélène and Chaté, Hugues and Theraulaz, Guy and Bon, Rémi},
  journal   = {Proceedings of the National Academy of Sciences},
  volume    = {112},
  number    = {41},
  pages     = {12729--12734},
  year      = {2015},
  publisher = {National Academy of Sciences},
  doi       = {10.1073/pnas.1503749112}
}

@article{Guillet2025Melting_of_nonreciprocal_solids,
  title={Melting of nonreciprocal solids: How dislocations propel and fission in flowing crystals},
  author={Guillet, St{\'e}phane and Poncet, Alexis and Le Blay, Marine and Irvine, William TM and Vitelli, Vincenzo and Bartolo, Denis},
  journal={Proceedings of the National Academy of Sciences},
  volume={122},
  number={15},
  pages={e2412993122},
  year={2025},
  publisher={National Academy of Sciences},
  url={https://www.pnas.org/doi/abs/10.1073/pnas.2412993122}
}

@article{Oza2019PRX_Lattices_Flapping_Swimmers,
  author    = {Anand U. Oza and Leif Ristroph and Michael J. Shelley},
  title     = {Lattices of Hydrodynamically Interacting Flapping Swimmers},
  journal   = {Physical Review X},
  volume    = {9},
  number    = {4},
  pages     = {041024},
  year      = {2019},
  doi       = {10.1103/PhysRevX.9.041024},
  url       = {https://doi.org/10.1103/PhysRevX.9.041024}
}

@article{hickey2023nonreciprocalcilium,
  title={Nonreciprocal interactions give rise to fast cilium synchronization in finite systems},
  author={Hickey, David J and Golestanian, Ramin and Vilfan, Andrej},
  journal={Proceedings of the National Academy of Sciences},
  volume={120},
  number={40},
  pages={e2307279120},
  year={2023},
  publisher={National Academy of Sciences},
  url={https://www.pnas.org/doi/10.1073/pnas.2307279120}
}

@article{Filella2018PRL_Collective_Fish_Behavior_Hydrodynamic_Interactions,
  author    = {Audrey Filella and François Nadal and Clément Sire and Eva Kanso and Christophe Eloy},
  title     = {Model of Collective Fish Behavior with Hydrodynamic Interactions},
  journal   = {Physical Review Letters},
  volume    = {120},
  number    = {19},
  pages     = {198101},
  year      = {2018},
  doi       = {10.1103/PhysRevLett.120.198101},
  url       = {https://doi.org/10.1103/PhysRevLett.120.198101}
}

@article{BradyWu2025A_three_body_problem,
  title={Nonreciprocity and multibody interactions in acoustically levitated particle systems: A three-body problem},
  author={Wu, Brady and Mao, Qinghao and VanSaders, Bryan and Jaeger, Heinrich M},
  journal={Physical Review E},
  volume={112},
  number={3},
  pages={035410},
  year={2025},
  publisher={APS},
  url={https://journals.aps.org/pre/abstract/10.1103/tpzc-qms7}
}

@article{EllaKing2025Scattered_waves_fuel_emergent_activity,
  author    = {Ella M. King and Mia C. Morrell and Jacqueline B. Sustiel and Matthew Gronert and Hayden Pastor and David G. Grier},
  title     = {Scattered waves fuel emergent activity},
  journal   = {Physical Review Research},
  volume    = {7},
  number    = {1},
  pages     = {013055},
  year      = {2025},
  doi       = {10.1103/PhysRevResearch.7.013055},
  url       = {https://doi.org/10.1103/PhysRevResearch.7.013055}
}

@article{Morrell2025Nonreciprocal_wave-mediated_interactions_power_a_classical_time_crystal,
  title={Nonreciprocal wave-mediated interactions power a classical time crystal},
  author={Morrell, Mia C and Elliott, Leela and Grier, David G},
  journal={Physical Review Letters},
  volume={136},
  number={5},
  pages={057201},
  year={2026},
  publisher={APS},
  url= {https://journals.aps.org/prl/abstract/10.1103/zjzk-t81n}
}

@article{yu2025pnas,
  title={Physics-tailored machine learning reveals unexpected physics in dusty plasmas},
  author={Yu, Wentao and Abdelaleem, Eslam and Nemenman, Ilya and Burton, Justin C},
  journal={Proceedings of the National Academy of Sciences},
  volume={122},
  number={31},
  pages={e2505725122},
  year={2025},
  publisher={National Academy of Sciences},
  url = {https://www.pnas.org/doi/abs/10.1073/pnas.2505725122}
}

@article{efi2019self,
  title={Self-driven fractional rotational diffusion of the harmonic three-mass system},
  author={Saporta Katz, Ori and Efrati, Efi},
  journal={Physical review letters},
  volume={122},
  number={2},
  pages={024102},
  year={2019},
  publisher={APS},
  url = {https://journals.aps.org/prl/abstract/10.1103/PhysRevLett.122.024102}
}

@article{kovacic2018mathieu,
  title={Mathieu's equation and its generalizations: overview of stability charts and their features},
  author={Kovacic, Ivana and Rand, Richard and Mohamed Sah, Si},
  journal={Applied Mechanics Reviews},
  volume={70},
  number={2},
  pages={020802},
  year={2018},
  publisher={American Society of Mechanical Engineers},
  url={https://asmedigitalcollection.asme.org/appliedmechanicsreviews/article-abstract/70/2/020802/368765/Mathieu-s-Equation-and-Its-Generalizations}
}

@article{miles1990parametrically,
  title={Parametrically forced surface waves},
  author={Miles, John and Henderson, Diane},
  journal={Annual Review of Fluid Mechanics},
  volume={22},
  number={1},
  pages={143--165},
  year={1990},
  url={https://www.annualreviews.org/content/journals/10.1146/annurev.fl.22.010190.001043}
}

@article{mao2025structural,
  title={Structural reconfiguration of interacting multi-particle systems through parametric pumping},
  author={Mao, Qinghao and Wu, Brady and VanSaders, Bryan and Jaeger, Heinrich M},
  journal={Nature communications},
  volume={16},
  number={1},
  pages={4637},
  year={2025},
  publisher={Nature Publishing Group UK London},
  url={https://www.nature.com/articles/s41467-025-59631-3}
}

@article{nunomura1999delaycharginginstability,
  title={Instability of dust particles in a coulomb crystal due to delayed charging},
  author={Nunomura, S and Misawa, T and Ohno, N and Takamura, S},
  journal={Physical review letters},
  volume={83},
  number={10},
  pages={1970},
  year={1999},
  publisher={APS},
  url={https://journals.aps.org/prl/abstract/10.1103/PhysRevLett.83.1970}
}

@article{ivlev2000influence,
  title={Influence of charge variation on particle oscillations in the plasma sheath},
  author={Ivlev, AV and Konopka, U and Morfill, G},
  journal={Physical Review E},
  volume={62},
  number={2},
  pages={2739},
  year={2000},
  publisher={APS},
  url={https://journals.aps.org/pre/abstract/10.1103/PhysRevE.62.2739}
}

@article{mendez2020origin,
  title={Origin of large-amplitude oscillations of dust particles in a plasma sheath},
  author={M{\'e}ndez Harper, Joshua and Gogia, Guram and Wu, Brady and Laseter, Zachary and Burton, Justin C},
  journal={Physical Review Research},
  volume={2},
  number={3},
  pages={033500},
  year={2020},
  publisher={APS},
  url={https://journals.aps.org/prresearch/abstract/10.1103/PhysRevResearch.2.033500}
}

@article{douglass2012determination,
  title={Determination of the levitation limits of dust particles within the sheath in complex plasma experiments},
  author={Douglass, Angela and Land, Victor and Qiao, Ke and Matthews, Lorin and Hyde, Truell},
  journal={Physics of Plasmas},
  volume={19},
  number={1},
  year={2012},
  publisher={AIP Publishing},
  url={https://pubs.aip.org/aip/pop/article-abstract/19/1/013707/107496/Determination-of-the-levitation-limits-of-dust?redirectedFrom=fulltext}
}

@article{lorin2019nonlinear,
  title={Nonlinear mode coupling and internal resonance observed in a dusty plasma},
  author={Ding, Zhiyue and Qiao, Ke and Ernst, Nicholas and Kong, Jie and Chen, Mudi and Matthews, Lorin S and Hyde, Truell W},
  journal={New Journal of Physics},
  volume={21},
  number={10},
  pages={103051},
  year={2019},
  publisher={IOP Publishing},
  url={https://iopscience.iop.org/article/10.1088/1367-2630/ab4d95/meta}
}

@article{ivlev2015statistical,
  title={Statistical mechanics where Newton’s third law is broken},
  author={Ivlev, Alexei V and Bartnick, J{\"o}rg and Heinen, Marco and Du, C-R and Nosenko, V and L{\"o}wen, Hartmut},
  journal={Physical Review X},
  volume={5},
  number={1},
  pages={011035},
  year={2015},
  publisher={APS},
  url={https://journals.aps.org/prx/abstract/10.1103/PhysRevX.5.011035}
}

@article{nadolny2025Nonreciprocal_synchronization_of_active_quantum_spins,
  title={Nonreciprocal synchronization of active quantum spins},
  author={Nadolny, Tobias and Bruder, Christoph and Brunelli, Matteo},
  journal={Physical Review X},
  volume={15},
  number={1},
  pages={011010},
  year={2025},
  publisher={APS},
  url={https://journals.aps.org/prx/abstract/10.1103/PhysRevX.15.011010}
}

@article{avni2025nonreciprocal_ising,
  title={Nonreciprocal ising model},
  author={Avni, Yael and Fruchart, Michel and Martin, David and Seara, Daniel and Vitelli, Vincenzo},
  journal={Physical Review Letters},
  volume={134},
  number={11},
  pages={117103},
  year={2025},
  publisher={APS},
  url={https://journals.aps.org/prl/abstract/10.1103/PhysRevLett.134.117103}
}

@article{brauns2024nonreciprocal_pattern_formation,
  title={Nonreciprocal pattern formation of conserved fields},
  author={Brauns, Fridtjof and Marchetti, M Cristina},
  journal={Physical Review X},
  volume={14},
  number={2},
  pages={021014},
  year={2024},
  publisher={APS},
  url={https://journals.aps.org/prx/abstract/10.1103/PhysRevX.14.021014}
}

@article{fruchart2021Non-reciprocal-phase-transitions,
  title={Non-reciprocal phase transitions},
  author={Fruchart, Michel and Hanai, Ryo and Littlewood, Peter B and Vitelli, Vincenzo},
  journal={Nature},
  volume={592},
  number={7854},
  pages={363--369},
  year={2021},
  publisher={Nature Publishing Group UK London},
  url={https://www.nature.com/articles/s41586-021-03375-9}
}

@article{martin2025Transition-to-Collective-Motion-in-Nonreciprocal-Active-Matter,
  title={Transition to Collective Motion in Nonreciprocal Active Matter: Coarse Graining Agent-Based Models into Fluctuating Hydrodynamics},
  author={Martin, David and Seara, Daniel and Avni, Yael and Fruchart, Michel and Vitelli, Vincenzo},
  journal={Physical Review X},
  volume={15},
  number={4},
  pages={041015},
  year={2025},
  publisher={APS},
  url={https://journals.aps.org/prx/abstract/10.1103/PhysRevX.15.041015}
}

@article{tan2022odd-living-chiral-crystals,
  title={Odd dynamics of living chiral crystals},
  author={Tan, Tzer Han and Mietke, Alexander and Li, Junang and Chen, Yuchao and Higinbotham, Hugh and Foster, Peter J and Gokhale, Shreyas and Dunkel, J{\"o}rn and Fakhri, Nikta},
  journal={Nature},
  volume={607},
  number={7918},
  pages={287--293},
  year={2022},
  publisher={Nature Publishing Group UK London},
  url={https://www.nature.com/articles/s41586-022-04889-6}
}

@article{battle2016broken-detailed-balance,
  title={Broken detailed balance at mesoscopic scales in active biological systems},
  author={Battle, Christopher and Broedersz, Chase P and Fakhri, Nikta and Geyer, Veikko F and Howard, Jonathon and Schmidt, Christoph F and MacKintosh, Fred C},
  journal={Science},
  volume={352},
  number={6285},
  pages={604--607},
  year={2016},
  publisher={American Association for the Advancement of Science},
  url={https://www.science.org/doi/abs/10.1126/science.aac8167}
}

@article{cavagna2010scale-Scale-free-correlations-flocks,
  title={Scale-free correlations in starling flocks},
  author={Cavagna, Andrea and Cimarelli, Alessio and Giardina, Irene and Parisi, Giorgio and Santagati, Raffaele and Stefanini, Fabio and Viale, Massimiliano},
  journal={Proceedings of the National Academy of Sciences},
  volume={107},
  number={26},
  pages={11865--11870},
  year={2010},
  publisher={National Academy of Sciences},
  url={https://www.pnas.org/doi/10.1073/pnas.1005766107}
}

@article{ballerini2008flock-animal-collective-behavior-depends-on-topological-rather-than-metric-distance,
  title={Interaction ruling animal collective behavior depends on topological rather than metric distance: Evidence from a field study},
  author={Ballerini, Michele and Cabibbo, Nicola and Candelier, Raphael and Cavagna, Andrea and Cisbani, Evaristo and Giardina, Irene and Lecomte, Vivien and Orlandi, Alberto and Parisi, Giorgio and Procaccini, Andrea and others},
  journal={Proceedings of the national academy of sciences},
  volume={105},
  number={4},
  pages={1232--1237},
  year={2008},
  publisher={National Academy of Sciences},
  url={https://www.pnas.org/doi/abs/10.1073/pnas.0711437105}
}

@article{nagy2010hierarchical-group-dynamics-in-pigeon-flocks,
  title={Hierarchical group dynamics in pigeon flocks},
  author={Nagy, M{\'a}t{\'e} and {\'A}kos, Zsuzsa and Biro, Dora and Vicsek, Tam{\'a}s},
  journal={Nature},
  volume={464},
  number={7290},
  pages={890--893},
  year={2010},
  publisher={Nature Publishing Group UK London},
  url={https://www.nature.com/articles/nature08891}
}

@article{guga2017emergent,
  title={Emergent bistability and switching in a nonequilibrium crystal},
  author={Gogia, Guram and Burton, Justin C},
  journal={Physical Review Letters},
  volume={119},
  number={17},
  pages={178004},
  year={2017},
  publisher={APS},
  url={https://journals.aps.org/prl/abstract/10.1103/PhysRevLett.119.178004}
}

@article{wu2023hydrodynamic,
  title={Hydrodynamic coupling melts acoustically levitated crystalline rafts},
  author={Wu, Brady and VanSaders, Bryan and Lim, Melody X and Jaeger, Heinrich M},
  journal={Proceedings of the National Academy of Sciences},
  volume={120},
  number={29},
  pages={e2301625120},
  year={2023},
  publisher={National Academy of Sciences}
}

@article{guga2020intermittent,
  title={Intermittent turbulence in a many-body system},
  author={Gogia, Guram and Yu, Wentao and Burton, Justin C},
  journal={Physical Review Research},
  volume={2},
  number={2},
  pages={023250},
  year={2020},
  publisher={APS},
  url={https://journals.aps.org/prresearch/abstract/10.1103/PhysRevResearch.2.023250}
}

@article{Active-particles-rmp,
  title={Active particles in complex and crowded environments},
  author={Bechinger, Clemens and Di Leonardo, Roberto and L{\"o}wen, Hartmut and Reichhardt, Charles and Volpe, Giorgio and Volpe, Giovanni},
  journal={Reviews of modern physics},
  volume={88},
  number={4},
  pages={045006},
  year={2016},
  publisher={APS},
  url={https://journals.aps.org/rmp/abstract/10.1103/RevModPhys.88.045006}
}

@article{you2020nonreciprocity-traveling-states,
  title={Nonreciprocity as a generic route to traveling states},
  author={You, Zhihong and Baskaran, Aparna and Marchetti, M Cristina},
  journal={Proceedings of the National Academy of Sciences},
  volume={117},
  number={33},
  pages={19767--19772},
  year={2020},
  publisher={National Academy of Sciences},
  url={https://www.pnas.org/doi/abs/10.1073/pnas.2010318117}
}

@article{saha2020nonreciprocal-Cahn-Hilliard-model,
  title={Scalar active mixtures: The nonreciprocal Cahn-Hilliard model},
  author={Saha, Suropriya and Agudo-Canalejo, Jaime and Golestanian, Ramin},
  journal={Physical Review X},
  volume={10},
  number={4},
  pages={041009},
  year={2020},
  publisher={APS},
  url={https://journals.aps.org/prx/abstract/10.1103/PhysRevX.10.041009}
}

@article{scheibner2020-Odd-elasticity,
  title={Odd elasticity},
  author={Scheibner, Colin and Souslov, Anton and Banerjee, Debarghya and Sur{\'o}wka, Piotr and Irvine, William TM and Vitelli, Vincenzo},
  journal={Nature Physics},
  volume={16},
  number={4},
  pages={475--480},
  year={2020},
  publisher={Nature Publishing Group UK London},
  url={https://www.nature.com/articles/s41567-020-0795-y}
}

@article{baconnier2022Selective-and-collective-actuation-in-active-solids,
  title={Selective and collective actuation in active solids},
  author={Baconnier, Paul and Shohat, Dor and L{\'o}pez, C Hern{\'a}ndez and Coulais, Corentin and D{\'e}mery, Vincent and D{\"u}ring, Gustavo and Dauchot, Olivier},
  journal={Nature Physics},
  volume={18},
  number={10},
  pages={1234--1239},
  year={2022},
  publisher={Nature Publishing Group UK London},
  url={https://www.nature.com/articles/s41567-022-01704-x}
}

@article{veenstra2025adaptive-locomotion-of-active-solids,
  title={Adaptive locomotion of active solids},
  author={Veenstra, Jonas and Scheibner, Colin and Brandenbourger, Martin and Binysh, Jack and Souslov, Anton and Vitelli, Vincenzo and Coulais, Corentin},
  journal={Nature},
  volume={639},
  number={8056},
  pages={935--941},
  year={2025},
  publisher={Nature Publishing Group UK London},
  url={https://www.nature.com/articles/s41586-025-08646-3}
}

@article{palacci2013Living-crystals-of-light-activated-colloidal-surfers,
  title={Living crystals of light-activated colloidal surfers},
  author={Palacci, Jeremie and Sacanna, Stefano and Steinberg, Asher Preska and Pine, David J and Chaikin, Paul M},
  journal={Science},
  volume={339},
  number={6122},
  pages={936--940},
  year={2013},
  publisher={American Association for the Advancement of Science},
  url={https://www.science.org/doi/abs/10.1126/science.1230020}
}

@article{aubret2018targeted-assembly-and-synchronization-of-self-spinning-microgears,
  title={Targeted assembly and synchronization of self-spinning microgears},
  author={Aubret, Antoine and Youssef, Mena and Sacanna, Stefano and Palacci, J{\'e}r{\'e}mie},
  journal={Nature Physics},
  volume={14},
  number={11},
  pages={1114--1118},
  year={2018},
  publisher={Nature Publishing Group UK London},
  url={https://www.nature.com/articles/s41567-018-0227-4}
}

@article{osat2023non-reciprocal-multifarious-dself-organization,
  title={Non-reciprocal multifarious self-organization},
  author={Osat, Saeed and Golestanian, Ramin},
  journal={Nature Nanotechnology},
  volume={18},
  number={1},
  pages={79--85},
  year={2023},
  publisher={Nature Publishing Group UK London},
  url={https://www.nature.com/articles/s41565-022-01258-2}
}

@article{osat2024escaping-kinetic-traps-using-nonreciprocal-interactions,
  title={Escaping kinetic traps using nonreciprocal interactions},
  author={Osat, Saeed and Metson, Jakob and Kardar, Mehran and Golestanian, Ramin},
  journal={Physical Review Letters},
  volume={133},
  number={2},
  pages={028301},
  year={2024},
  publisher={APS},
  url={https://journals.aps.org/prl/abstract/10.1103/PhysRevLett.133.028301}
}

@article{maity2023spontaneous-demixing-of-binary-colloidal-flocks,
  title={Spontaneous demixing of binary colloidal flocks},
  author={Maity, Samadarshi and Morin, Alexandre},
  journal={Physical Review Letters},
  volume={131},
  number={17},
  pages={178304},
  year={2023},
  publisher={APS},
  url={https://journals.aps.org/prl/abstract/10.1103/PhysRevLett.131.178304}
}

@article{berg1972chemotaxis-in-Escherichia-coli-analysed-by-three-dimensional-tracking,
  title={Chemotaxis in Escherichia coli analysed by three-dimensional tracking},
  author={Berg, Howard C and Brown, Douglas A},
  journal={Nature},
  volume={239},
  number={5374},
  pages={500--504},
  year={1972},
  publisher={Nature Publishing Group UK London},
  url={https://www.nature.com/articles/239500a0}
}

@article{brandenbourger2019Non-reciprocal-robotic-metamaterials,
  title={Non-reciprocal robotic metamaterials},
  author={Brandenbourger, Martin and Locsin, Xander and Lerner, Edan and Coulais, Corentin},
  journal={Nature communications},
  volume={10},
  number={1},
  pages={4608},
  year={2019},
  publisher={Nature Publishing Group UK London},
  url={https://www.nature.com/articles/s41467-019-12599-3}
}

@article{veenstra2024Non-reciprocal-topological-solitons,
  title={Non-reciprocal topological solitons in active metamaterials},
  author={Veenstra, Jonas and Gamayun, Oleksandr and Guo, Xiaofei and Sarvi, Anahita and Meinersen, Chris Ventura and Coulais, Corentin},
  journal={Nature},
  volume={627},
  number={8004},
  pages={528--533},
  year={2024},
  publisher={Nature Publishing Group UK London},
  url={https://www.nature.com/articles/s41586-024-07097-6}
}

@article{veenstra2025nonreciprocal-breathing-solitons,
  title={Nonreciprocal breathing solitons},
  author={Veenstra, Jonas and Gamayun, Oleksandr and Brandenbourger, Martin and van Gorp, Freek and Terwisscha-Dekker, Hans and Caux, Jean-S{\'e}bastien and Coulais, Corentin},
  journal={Physical Review X},
  volume={15},
  number={3},
  pages={031045},
  year={2025},
  publisher={APS},
  url={https://journals.aps.org/prx/abstract/10.1103/nrv2-9h8z}
}

@article{paxton2004catalytic-nanomotors,
  title={Catalytic nanomotors: autonomous movement of striped nanorods},
  author={Paxton, Walter F and Kistler, Kevin C and Olmeda, Christine C and Sen, Ayusman and St. Angelo, Sarah K and Cao, Yanyan and Mallouk, Thomas E and Lammert, Paul E and Crespi, Vincent H},
  journal={Journal of the American Chemical Society},
  volume={126},
  number={41},
  pages={13424--13431},
  year={2004},
  publisher={ACS Publications},
  url={https://pubs.acs.org/doi/10.1021/ja047697z}
}

@article{golestanian2005propulsion,
  title={Propulsion of a molecular machine by asymmetric distribution of reaction products},
  author={Golestanian, Ramin and Liverpool, Tanniemola B and Ajdari, Armand},
  journal={Physical review letters},
  volume={94},
  number={22},
  pages={220801},
  year={2005},
  publisher={APS},
  url={https://journals.aps.org/prl/abstract/10.1103/PhysRevLett.94.220801}
}

@article{schliwa2003molecular,
  title={Molecular motors},
  author={Schliwa, Manfred and Woehlke, G{\"u}nther},
  journal={Nature},
  volume={422},
  number={6933},
  pages={759--765},
  year={2003},
  publisher={Nature Publishing Group UK London},
  url={https://www.nature.com/articles/nature01601}
}

@article{julicher1997modeling,
  title={Modeling molecular motors},
  author={J{\"u}licher, Frank and Ajdari, Armand and Prost, Jacques},
  journal={Reviews of Modern Physics},
  volume={69},
  number={4},
  pages={1269},
  year={1997},
  publisher={APS},
  url={https://journals.aps.org/rmp/abstract/10.1103/RevModPhys.69.1269}
}

@article{loos2020irreversibility,
  title={Irreversibility, heat and information flows induced by non-reciprocal interactions},
  author={Loos, Sarah AM and Klapp, Sabine HL},
  journal={New Journal of Physics},
  volume={22},
  number={12},
  pages={123051},
  year={2020},
  publisher={IOP Publishing},
  url={https://iopscience.iop.org/article/10.1088/1367-2630/abcc1e/meta}
}

@article{zhang2023entropy,
  title={Entropy production of nonreciprocal interactions},
  author={Zhang, Ziluo and Garcia-Millan, Rosalba},
  journal={Physical Review Research},
  volume={5},
  number={2},
  pages={L022033},
  year={2023},
  publisher={APS},
  url={https://journals.aps.org/prresearch/abstract/10.1103/PhysRevResearch.5.L022033}
}

@article{ai2023brownian,
  title={Brownian motors powered by nonreciprocal interactions},
  author={Ai, Bao-quan},
  journal={Physical Review E},
  volume={108},
  number={6},
  pages={064409},
  year={2023},
  publisher={APS},
  url={https://journals.aps.org/pre/abstract/10.1103/PhysRevE.108.064409}
}

@article{thomas1994plasma,
  title={Plasma crystal: Coulomb crystallization in a dusty plasma},
  author={Thomas, H and Morfill, GE and Demmel, V and Goree, J and Feuerbacher, B and M{\"o}hlmann, D},
  journal={Physical Review Letters},
  volume={73},
  number={5},
  pages={652},
  year={1994},
  publisher={APS},
  url={https://journals.aps.org/prl/abstract/10.1103/PhysRevLett.73.652}
}

@article{chu1994direct,
  title={Direct observation of Coulomb crystals and liquids in strongly coupled rf dusty plasmas},
  author={Chu, JH and Lin, I},
  journal={Physical review letters},
  volume={72},
  number={25},
  pages={4009},
  year={1994},
  publisher={APS},
  url={https://journals.aps.org/prl/abstract/10.1103/PhysRevLett.72.4009}
}

@article{juan1998observation,
  title={Observation of dust Coulomb clusters in a plasma trap},
  author={Juan, Wen-Tau and Huang, Zen-Hong and Hsu, Ju-Wang and Lai, Yin-Ju and others},
  journal={Physical Review E},
  volume={58},
  number={6},
  pages={R6947},
  year={1998},
  publisher={APS},
  url={https://journals.aps.org/pre/abstract/10.1103/PhysRevE.58.R6947}
}

@article{couedel2010direct,
  title={Direct observation of mode-coupling instability in two-dimensional plasma crystals},
  author={Cou{\"e}del, L and Nosenko, V and Ivlev, AV and Zhdanov, SK and Thomas, HM and Morfill, GE},
  journal={Physical review letters},
  volume={104},
  number={19},
  pages={195001},
  year={2010},
  publisher={APS},
  url={https://journals.aps.org/prl/abstract/10.1103/PhysRevLett.104.195001}
}

@article{liu2010mode,
  title={Mode coupling for phonons in a single-layer dusty plasma crystal},
  author={Liu, Bin and Goree, J and Feng, Yan},
  journal={Physical review letters},
  volume={105},
  number={8},
  pages={085004},
  year={2010},
  publisher={APS},
  url={https://journals.aps.org/prl/abstract/10.1103/PhysRevLett.105.085004}
}

@article{matthews2020dust,
  title={Dust charging in dynamic ion wakes},
  author={Matthews, Lorin Swint and Sanford, Dustin L and Kostadinova, Evdokiya G and Ashrafi, Khandaker Sharmin and Guay, Evelyn and Hyde, Truell W},
  journal={Physics of Plasmas},
  volume={27},
  number={2},
  year={2020},
  publisher={AIP Publishing},
  url={https://pubs.aip.org/aip/pop/article-abstract/27/2/023703/1062796/Dust-charging-in-dynamic-ion-wakes?redirectedFrom=fulltext}
}

@article{melzer1999transition,
  title={Transition from attractive to repulsive forces between dust molecules in a plasma sheath},
  author={Melzer, A and Schweigert, VA and Piel, A},
  journal={Physical review letters},
  volume={83},
  number={16},
  pages={3194},
  year={1999},
  publisher={APS},
  url={https://journals.aps.org/prl/abstract/10.1103/PhysRevLett.83.3194}
}

@article{hebner2003dynamic,
  title={Dynamic probe of dust wakefield interactions using constrained collisions},
  author={Hebner, Gregory Albert and Riley, ME and Marder, BM},
  journal={Physical Review E},
  volume={68},
  number={1},
  pages={016403},
  year={2003},
  publisher={APS},
  url={https://journals.aps.org/pre/abstract/10.1103/PhysRevE.68.016403}
}

@article{yu20233d,
  title={3D tracking of particles in a dusty plasma by laser sheet tomography},
  author={Yu, Wentao and Burton, Justin C},
  journal={Physics of Plasmas},
  volume={30},
  number={6},
  year={2023},
  publisher={AIP Publishing},
  url={https://pubs.aip.org/aip/pop/article-abstract/30/6/063701/2896141/3D-tracking-of-particles-in-a-dusty-plasma-by?redirectedFrom=fulltext}
}

@software{Allan2021trackpy,
  author       = {Allan, D. B. and Caswell, T. and Keim, N. C. and van der Wel, C. M. and Verweij, R. W.},
  title        = {soft-matter/trackpy: Trackpy v0.5.0 (v0.5.0)},
  year         = {2021},
  publisher    = {Zenodo},
  doi          = {10.5281/zenodo.4682814},
  url          = {https://doi.org/10.5281/zenodo.4682814}
}

@article{ivanov2005melting,
  title={Melting dynamics of finite clusters in dusty plasmas},
  author={Ivanov, Yuriy and Melzer, Andre},
  journal={Physics of plasmas},
  volume={12},
  number={7},
  year={2005},
  publisher={AIP Publishing},
  url={https://pubs.aip.org/aip/pop/article-abstract/12/7/072110/262526/Melting-dynamics-of-finite-clusters-in-dusty?redirectedFrom=fulltext}
}

@article{iwai1987geometric,
  title={A geometric setting for internal motions of the quantum three-body system},
  author={Iwai, Toshihiro},
  journal={Journal of mathematical physics},
  volume={28},
  number={6},
  pages={1315--1326},
  year={1987},
  publisher={American Institute of Physics}
}

@article{qiao2013mode,
  title={Mode couplings and conversions for horizontal dust particle pairs in complex plasmas},
  author={Qiao, Ke and Kong, Jie and Zhang, Zhuanhao and Matthews, Lorin S and Hyde, Truell W},
  journal={IEEE Transactions on Plasma Science},
  volume={41},
  number={4},
  pages={745--753},
  year={2013},
  publisher={IEEE}
}

@article{yaroshenko2006vibrations,
  title={Vibrations of a pair microparticles suspended in a plasma sheath},
  author={Yaroshenko, VV and Vladimirov, SV and Morfill, GE},
  journal={New Journal of Physics},
  volume={8},
  number={9},
  pages={201--201},
  year={2006}
}

@article{jung2018experiments,
  title={Experiments on wake structures behind a microparticle in a magnetized plasma flow},
  author={Jung, Hendrik and Greiner, Franko and Piel, Alexander and Miloch, Wojciech J},
  journal={Physics of Plasmas},
  volume={25},
  number={7},
  year={2018},
  publisher={AIP Publishing}
}

@article{carstensen2012ion,
  title={Ion-wake-mediated particle interaction in a magnetized-plasma flow},
  author={Carstensen, Jan and Greiner, Franko and Piel, Alexander},
  journal={Physical Review Letters},
  volume={109},
  number={13},
  pages={135001},
  year={2012},
  publisher={APS}
}

@article{qiao2014mode,
  title={Mode coupling and resonance instabilities in quasi-two-dimensional dust clusters in complex plasmas},
  author={Qiao, Ke and Kong, Jie and Carmona-Reyes, Jorge and Matthews, Lorin S and Hyde, Truell W},
  journal={Physical Review E},
  volume={90},
  number={3},
  pages={033109},
  year={2014},
  publisher={APS}
}

@article{kompaneets2006dust,
  title={Dust clusters with non-Hamiltonian particle dynamics},
  author={Kompaneets, R and Vladimirov, SV and Ivlev, AV and Tsytovich, V and Morfill, G},
  journal={Physics of plasmas},
  volume={13},
  number={7},
  year={2006},
  publisher={AIP Publishing}
}

@inbook{brunton_kutz_2022_sec1_5,
  author    = {Brunton, Steven L. and Kutz, J. Nathan},
  title     = {Data-Driven Science and Engineering: Machine Learning, Dynamical Systems, and Control},
  edition   = {2},
  publisher = {Cambridge University Press},
  year      = {2022},
  chapter   = {1.5},
  doi       = {10.1017/9781009089517},
  note      = {Section 1.5}
}

@article{epstein1924resistance,
  title={On the resistance experienced by spheres in their motion through gases},
  author={Epstein, Paul S},
  journal={Physical Review},
  volume={23},
  number={6},
  pages={710},
  year={1924},
  publisher={APS},
  url={https://journals.aps.org/pr/abstract/10.1103/PhysRev.23.710}
}

@article{burov2017single,
  title={Single-pixel interior filling function approach for detecting and correcting errors in particle tracking},
  author={Burov, Stanislav and Figliozzi, Patrick and Lin, Binhua and Rice, Stuart A and Scherer, Norbert F and Dinner, Aaron R},
  journal={Proceedings of the National Academy of Sciences},
  volume={114},
  number={2},
  pages={221--226},
  year={2017},
  publisher={National Academy of Sciences},
  url={https://www.pnas.org/doi/10.1073/pnas.1619104114}
}

@article{ishihara1997wake,
  title={Wake potential of a dust grain in a plasma with ion flow},
  author={Ishihara, Osamu and Vladimirov, Sergey V},
  journal={Physics of Plasmas},
  volume={4},
  number={1},
  pages={69--74},
  year={1997},
  publisher={American Institute of Physics},
  url={https://pubs.aip.org/aip/pop/article-abstract/4/1/69/262520/Wake-potential-of-a-dust-grain-in-a-plasma-with?redirectedFrom=PDF}
}

@article{vladimirov1995attraction,
  title={Attraction of charged particulates in plasmas with finite flows},
  author={Vladimirov, Sergey V and Nambu, Mitsuhiro},
  journal={Physical Review E},
  volume={52},
  number={3},
  pages={R2172},
  year={1995},
  publisher={APS},
  url={https://journals.aps.org/pre/abstract/10.1103/PhysRevE.52.R2172}
}

@article{lampe2000interactions,
  title={Interactions between dust grains in a dusty plasma},
  author={Lampe, Martin and Joyce, Glenn and Ganguli, Gurudas and Gavrishchaka, Valeriy},
  journal={Physics of Plasmas},
  volume={7},
  number={10},
  pages={3851--3861},
  year={2000},
  publisher={American Institute of Physics},
  url={https://pubs.aip.org/aip/pop/article-abstract/7/10/3851/264667/Interactions-between-dust-grains-in-a-dusty-plasma?redirectedFrom=fulltext},
  url={https://pubs.aip.org/aip/pop/article-abstract/7/10/3851/264667/Interactions-between-dust-grains-in-a-dusty-plasma}
}

@article{parker2025symmetry,
  title={Symmetry breaking-induced N-body electrodynamic forces in optical matter systems},
  author={Parker, John and Nagasamudram, Spoorthi and Peterson, Curtis and Li, Yanzeng and Soleimanikahnoj, Sina and Rice, Stuart A and Scherer, Norbert F},
  journal={Nature Communications},
  volume={16},
  number={1},
  pages={6294},
  year={2025},
  publisher={Nature Publishing Group UK London},
  url={https://www.nature.com/articles/s41467-025-61616-1}
}

@article{chen2021data,
  title={Data-driven reaction coordinate discovery in overdamped and non-conservative systems: application to optical matter structural isomerization},
  author={Chen, Shiqi and Peterson, Curtis W and Parker, John A and Rice, Stuart A and Ferguson, Andrew L and Scherer, Norbert F},
  journal={Nature Communications},
  volume={12},
  number={1},
  pages={2548},
  year={2021},
  publisher={Nature Publishing Group UK London},
  url={https://www.nature.com/articles/s41467-021-22794-w}
}

@article{yaroshenko2002parametric,
  title={Parametric excitation of low frequency waves in complex (dusty) plasmas},
  author={Yaroshenko, V and Morfill, GE},
  journal={Physics of Plasmas},
  volume={9},
  number={11},
  pages={4495--4499},
  year={2002},
  url={https://pubs.aip.org/aip/pop/article-abstract/9/11/4495/451533/Parametric-excitation-of-low-frequency-waves-in?redirectedFrom=fulltext}
}

@article{shi2025electrostatics,
  title={Electrostatics overcome acoustic collapse to assemble, adapt, and activate levitated matter},
  author={Shi, Sue and H{\"u}bl, Maximilian C and Grosjean, Galien and Goodrich, Carl P and Waitukaitis, Scott},
  journal={Proceedings of the National Academy of Sciences},
  volume={122},
  number={50},
  pages={e2516865122},
  year={2025},
  publisher={National Academy of Sciences},
  url={https://www.pnas.org/doi/abs/10.1073/pnas.2516865122}
}

\newpage




\newpage

\newpage

\newpage

\newpage 

\newpage

\newpage 

\newpage

\newpage

\newpage

\bigskip

\newpage

\end{document}